\documentclass{article}
\usepackage{url}
\usepackage{amsmath}
\usepackage{amsfonts}
\usepackage{graphicx} 
\usepackage{arxiv_new}
\usepackage{nicefrac}       
\usepackage{microtype}      
\usepackage{lipsum}
\usepackage{hyperref}
\usepackage{mathtools}
\usepackage{subcaption}
\usepackage{xcolor} 
\newcommand\blfootnote[1]{
    \begingroup
    \renewcommand\thefootnote{}\footnote{#1}
    \addtocounter{footnote}{-1}
    \endgroup
}

\title{Kilometer-Scale Convection Allowing Model Emulation using Generative Diffusion Modeling}

\author{
Jaideep Pathak\textsuperscript{*}\textsuperscript{1},
Yair Cohen\textsuperscript{*}\textsuperscript{1},
Piyush Garg\textsuperscript{*}\textsuperscript{1},
Peter Harrington\textsuperscript{*}\textsuperscript{2},
Noah Brenowitz\textsuperscript{1},
Dale Durran\textsuperscript{1,3},\\
Morteza Mardani\textsuperscript{1},
Arash Vahdat\textsuperscript{1},
Shaoming Xu\textsuperscript{\textdagger}\textsuperscript{4},
Karthik Kashinath\textsuperscript{1},
Michael Pritchard\textsuperscript{1}
}
\date{\today}

\begin{document}

\maketitle

\begin{center}
\textsuperscript{1}NVIDIA Corporation \\
\textsuperscript{2}Lawrence Berkeley National Laboratory\\
\textsuperscript{3}University of Washington \\
\textsuperscript{4}University of Minnesota \\
\textsuperscript{*}Equal Contribution
\end{center}

\vspace{0.5in}

\begin{abstract}
Storm-scale convection-allowing models (CAMs) are an important tool for predicting the evolution of thunderstorms and mesoscale convective systems that result in damaging extreme weather. By explicitly resolving convective dynamics within the atmosphere they afford meteorologists the nuance needed to provide outlook on hazard. Deep learning models have thus far not proven skilful at km-scale atmospheric simulation, despite being competitive at coarser resolution with state-of-the-art global, medium-range weather forecasting. We present a generative diffusion model called StormCast, which emulates the high-resolution rapid refresh (HRRR) model—NOAA's state-of-the-art 3km operational CAM. StormCast autoregressively predicts 99 state variables at km scale using a 1-hour time step, with dense vertical resolution in the atmospheric boundary layer, conditioned on 26 synoptic variables. We present evidence of successfully learnt km-scale dynamics including competitive 1-6 hour forecast skill for composite radar reflectivity alongside physically realistic convective cluster evolution, moist updrafts, and cold pool morphology. StormCast predictions maintain realistic power spectra for multiple predicted variables across multi-hour forecasts. Together, these results establish the potential for autoregressive ML to emulate CAMs -- opening up new km-scale frontiers for regional ML weather prediction and future climate hazard dynamical downscaling.

\end{abstract}

\section{Introduction}
Forecasting km-scale weather provides crucial information about extreme weather phenomena. At these scales, convection organizes into coherent systems such as mesoscale convective systems (MCSs), supercells, squall-lines, derechos and hurricanes~\cite{moncrieff2010multiscale, markowski2011mesoscale}, causing extreme winds, flooding, hailstorms, and tornadoes that can surpass annual damages of 50 billion US dollars in the United States alone \cite{NOAA2024}.\blfootnote{\textsuperscript{\textdagger} S. Xu's work was supported by an internship at Lawrence Berkeley National Laboratory.}  

Reliable numerical weather prediction of mesoscale phenomena, even at regional domains, incurs substantial computational costs due to the km-scale spatial resolution needed to represent the underlying fluid-dynamic motions~\cite{weisman2008experiences, skamarock2018limited}. Several nations support such large modeling efforts, which when coupled with data assimilation can produce skilful forecasts of mesoscale weather~\cite{chen2020improving, dowell2022high, james2022high}. The massive computational expense of these regional models-- often referred to as convection-allowing models (CAMs) or convection-permitting models-- prohibits their adoption in many countries, while forcing a tradeoff between the resolution and the ensemble size available for probabilistic prediction for those that can afford it. The latency of producing, post-processing and serving such km-scale forecasts can prohibit the potential benefits of these modeling systems in early warning of severe storms, which are provided on the order of hours ahead of a coming event. Recent commentary~\cite{bauer2024if} highlights the challenge~\cite{schulthess2018reflecting} of serving ever higher resolution ensemble numerical weather forecasts~\cite{bauer2015quiet} in an operational setting, where the forecasts must be generated faster than the weather unfolds within the constraints of computational scaling laws~\cite{shalf2020future}.

At lower resolutions, Machine Learning (ML) models trained on data assimilating global models have emerged as emulators of numerical weather models. These ML models, with a spatial resolution of about 30km and a typical temporal resolution of 6 hours, have obtained up to four orders of magnitude speedups while maintaining comparable skill~\cite{pathak2022fourcastnet,bonev2023spherical,lam2022graphcast, bi2023accurate, lang2024aifs, price2023gencast}. As such, they present a pathway to increasing the ensemble size of their counterpart, numerical, global weather models that are used for operational forecasts ~\cite{mahesh2024a,mahesh2024b}. 

If ML models can likewise be demonstrated to successfully emulate km-scale atmospheric dynamics, they could be used by operational centers to better serve time-critical high-resolution ensemble forecasts, economically. 

We posit that this task is not trivial. The two main challenges are the difference in the dominant forces in the fluid dynamics at km-scale, and the temporal resolution of available km-scale data. The dynamics of the atmosphere at the km-scale are far more complicated than at the 30km scale. 
At these lower resolutions, and especially when sampled at a 6-hour resolution, atmospheric dynamics are constrained by hydrostatic balance in the vertical, such that vertical motions are orders of magnitude weaker than horizontal motion, and are nearly in geostrophic balance at midlatitudes in the horizontal. It is thus akin to 2D turbulence, which on Earth has a predictability window of about 2 weeks. However, at the km scale atmospheric dynamics often deviates from these balances leading to strong, buoyancy-driven, convective motions. Thus, at the km scale, atmospheric motions are akin to 3D turbulence, which is inherently less predictable \cite{ooyama1982conceptual, raymond2015balanced}. Indeed, numerical modeling at the km scale is significantly harder than at 30km scale. Specifically, at these higher resolutions non-hydrostatic numerical dynamical cores are employed to capture the vertical accelerations in moist convection, vertical motions rival horizontal flows, and the subgrid physics packages in these models typically include many more species of precipitating hydrometeors and their interactions \cite{skamarock2018limited, skamarock2008description}.

The second challenge relates to an important observation that limits the size of deterministic ML model time steps.While not bound by constraints like the Courant-Friedrichs-Lewy (CFL) condition, deterministic ML models will predict the ensemble mean and filter-out any spatial scales not predictable at that lead time producing unrealistically smooth forecasts. Unfortunately, the most readily available and quality-controlled regional high-resolution and global low-resolution weather datasets are available at a one-hour sampling interval. At this time-scale only spatial scales larger than 10-20km are predictable \cite{Surcel2014-by,Surcel2015-qk} so deterministic models trained with such data will not produce realistic variability at the km-scale. Therefore the available data for regional convective-scale ML training is under-sampled in time compared to global reanalyses such as ERA5~\cite{hersbach2020era5}. Consistent with this view, Ref.~\cite{oskarsson2023graph} explored 10km resolution, regional determinstic ML emulation, resulting in forecasts with skill only for several smooth variables, but lacking the fine spatial scales for variables related to turbulence and storm formation processes, which are more stochastic. 

Generative ML models have shown promising results for km-scale stochastic fields in both downscaling (c.f. Ref~\cite{mardani2024residual, leinonen2020stochastic}) and nowcasting (c.f. Ref~\cite{ravuri2021skilful, leinonen2023latent}) as well as skilful ensemble forecasting at global 30km resolution~\cite{price2023gencast}. Some ML models have demonstrated skilful, high-resolution probabilistic outlooks summarizing statistical outcomes ~\cite{sonderby2020metnet, andrychowicz2023deep, espeholt2022deep} of precipitation and near-surface meteorological quantities, often directly from observational input data. Although such parsimonious ML models have shown remarkable abilities, the role of an operational CAM in meteorology extends beyond generating skilful quantitative estimates of precipitation and near-surface meteorological variables. CAMs are useful to meteorologists for tracking the evolution and structure of thunderstorms, monitoring the convective mode, and separating the stochastic from the deterministic evolution of convective organization across multiple interacting phenomena. For instance, estimating the likelihood of tornado formation tends to be based on knowledge of the predicted convective mode and storm structure~\cite{smith2012convective}. We hypothesize that predicting the complex evolution of storm-scale phenomena requires generative modeling of the full atmospheric state. 

These considerations motivate our work on realistic emulation of convective dynamics by directly predicting the temporal evolution of a dense km-scale state vector, and also distinguish it from prior works on regional forecasting, which do not provide a faithful representation of physical convective processes. In this work we present the first generative ML emulator of km-scale atmospheric dynamics named StormCast. StormCast is trained from an operational, radar-assimilating km-scale weather model that resolves a large range of convective motions. To capture the stochastic nature of the atmosphere at these scales we use a generative diffusion model~\cite{sohl2015deep, ho2020denoising, nichol2021improved, song2020score} using an optimized training and sampling framework presented in Ref.~\cite{karras2022elucidating} and following the two-step approach proposed in Ref~\cite{mardani2024residual}.

Our key results show that skilful emulation of CAMs is possible using generative diffusion. Its ease of inference allows us to generate ensembles, and produce Probability Matched Mean (PMM) of StormCast ensembles. Both Fractions Skill Score (FSS) and Root Mean Square Error (RMSE) of StormCast-PMM for composite radar reflectivity show comparable or even higher skill than HRRR forecasts at lead-times of up to 6 hours, at thresholds of 20dBZ, 30dBZ and 40dBZ indicating light, light-to-moderate and moderate rainfall. StormCast outputs also exhibit physically consistent convective dynamics. 
Moreover, StormCast produces vertical structure at the km-scale that is consistent with the underlying physics for convective motions and implies that a latent representation of cloud processes is required for generating plausible moist updrafts and cold pools. 

\section{Methods}
\subsection{Model Architecture: Generative Time-Stepped Diffusion Modeling}\label{sec:prob_forecast}
Consider the coarse-resolution synoptic-scale state \( \{S_t\}_{t \geq t_0} \) at 30km and the initial high-resolution mesoscale condition \( M_{t=t_0} \) at 3km. Our goal is to forecast the high-resolution mesoscale states \( M_t \) for \( t > t_0 \). At each time \( t \), we have access to the coarse-resolution forecast \( S_t \), typically available from another physics-based (e.g., IFS, GFS) or AI-based model (e.g., FourCastNet). Thus, we are motivated to learn the time-stepping distribution for time-ahead as the conditional distribution \( p_{\theta}(M_{t+1} \mid S_t, M_t) \). Once one has access to \( p_{\theta}(M_{t+1} \mid S_t, M_t) \), one can autoregressively sample the \( k \)-hour-ahead future for arbitrary \( k \) based on the initial condition \( M_{t_0} \) and the coarse-resolution forecast \( S_t \).

To learn the conditional distribution \( p_{\theta}(M_{t+1} \mid S_t, M_t) \), we adopt denoising diffusion models due to their effective mode coverage and training stability. Specifically, we follow the approach proposed in \cite{mardani2024residual} for learning the conditional distribution of weather data with both deterministic and stochastic dynamics. Following CorrDiff \cite{mardani2024residual}, we decompose the learning of \( p_{\theta} \) into two phases, namely deterministic regression, and stochastic diffusion:

\noindent\textbf{Phase 1: Deterministic Regression.}~First we learn a deterministic regression $F_\theta (M_t, S_t)$ to estimate the conditional mean, namely 
\begin{align}\label{eq:deterministic}
F_\theta (M_t, S_t) = \mathbb{E}[M_{t+1} \mid M_t, S_t] .
\end{align}
To do so, we can train a UNet based on the paired data samples $\lbrace (M_{t_i + 1}, M_{t_i}, S_{t_i} ) \rbrace_{i=0}^N$, and the MMSE criterion as follows
\begin{align}\label{eq:opt}
\theta^* = \underset{\theta}{\operatorname{argmin}} \; \frac{1}{N} \sum_{i=1}^N || M_{t_i+1} - F_\theta (M_{t_i}, S_{t_i}) ||_2^2 
\end{align}
With the learned regression model at hand, we can form the residuals
\begin{align}\label{eq:residual}
r_{t+1} \coloneq M_{t+1} - \mu_{t+1} \quad \text{where} \quad \mu_{t+1} := F_{\theta^*} (M_t, S_t).
\end{align}

\begin{figure}
    \centering
    \includegraphics[width=0.8\textwidth]{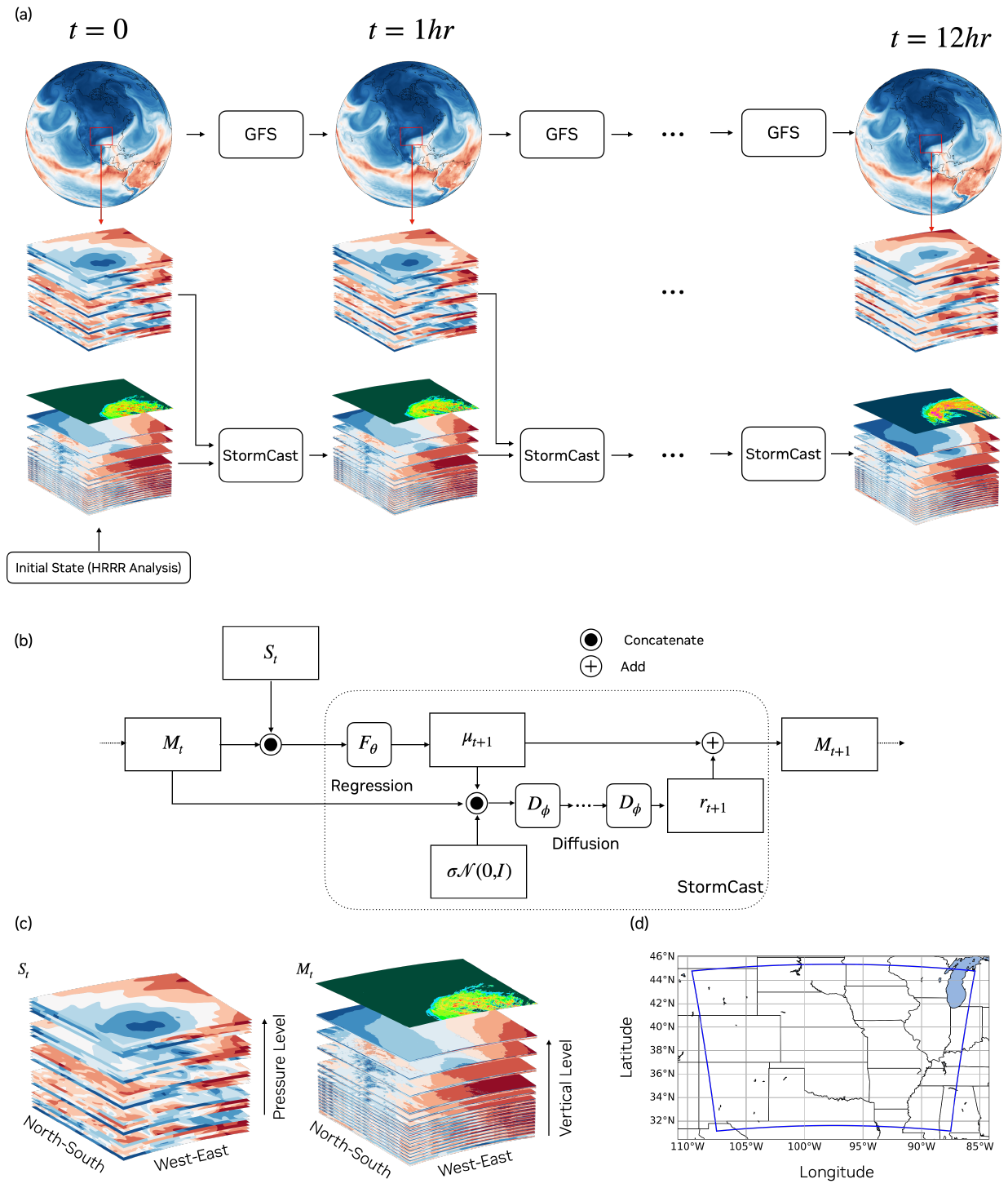}
    \caption{(a) StormCast generates an autoregressive forecast starting from an initial condition generated by HRRR analysis and using an hourly synoptic scale forecast produced by the GFS model. (b) Illustration of how StormCast generates a km-scale forecast in a two-step process. The synoptic-scale fields $S_t$ and the mesoscale fields $M_t$ at time $t$ are used to generate a one-hour ahead deterministic mean forecast $\mu_{t+1}$ using a neural network $F_\theta$ with a UNet architecture. The mean forecast $\mu_{t+1}$ and the $M_t$ are concatenated with a latent random gaussian noise vector and passed through a denoising diffusion model $D_\phi$ for a series of diffusion steps to sample an estimate of the residual forecast $r_{t+1}$ which is added to the mean $\mu_{t+1}$ to generate $M_{t+1}$ from the forecast distribution at time $t+1$. This process is repeated auto-regressively to generate a 12-hour forecast. The synoptic-scale conditioning $S_t$ at each time step $t$ is provided to the model via a 25km global model -- the NCEP GFS model in forecasting mode and ERA5 reanalysis in our hindcast tests. Panel (c) illustrates the stacked channels representing the synoptic-scale state $S_t$ on a pressure-level vertical grid and interpolated to the km-scale domain as well as the mesoscale state $M_t$ on the native HRRR model hybrid vertical grid. Refer to Tab.~\ref{tab:parameters} for the full channel set. The spatial extent of the domain of operation of the StormCast model is illustrated in panel (d) with a blue bounding box. The domain size is $1536$km $\times$ $1920$km }
    \label{fig:arch}
\end{figure}

\noindent\textbf{Phase 2: Stochastic Diffusion.} After forming the residuals, we train a diffusion model to learn the conditional distribution \( p(r_{t+1} \mid \mu_{t+1}, M_t, S_t) \). We use elucidated diffusion models (EDM)~\cite{Karras2022edm}, which are based on stochastic differential equations (SDEs) and offer a flexible way for auto-tuning. After training EDM, samples from the desired conditional distribution \( p(M_{t+1} \mid M_t, S_t) \) are obtained by combining the deterministic prediction and stochastic diffusion sample as \( r_{t+1} + \mu_{t+1} \). Figure~\ref{fig:arch}(b) illustrates the steps involved in generating a 1-hour forecast using StormCast. Figure~\ref{fig:arch}(a) shows how an autoregressive forecast is generated using a synoptic scale model and StormCast. Further details about the diffusion model configuration are presented in Appendix~\ref{app:diffusion}. 

\subsection{Training data}

\textbf{Mesoscale state:} Data for \( M_t \) come from the operational archive of the US km-scale forecasting model, the High-Resolution Rapid Refresh (HRRR) ~\cite{dowell2022high, james2022high} on its native model grid. To avoid non-stationarity from HRRR version updates, we only use data after July 2018 when HRRR v4 became operational. We use 6 dynamical variables (two horizontal wind components, temperature, geopotential height, pressure and specific humidity) from a subset of 16 of the available 50 hybrid vertical levels, retaining complete vertical sampling of the atmospheric boundary layer (see Table ~\ref{tab:parameters}). While we intended to train on the analysis state of HRRR, i.e. at the time of its hourly data assimilation, we inadvertently used data one hour subsequent, early in its forecast phase.

\textbf{Synoptic state (training):} Data for \( S_t \) come from the ERA5~\cite{hersbach2020era5} reanalysis. We use a sparse vertical sampling of 6 variables (two horizontal wind components, temperature, geopotential height, temperature, and specific humidity) across four pressure levels, each interpolated to the HRRR's native horizontal and vertical grid.  

\textbf{Temporal Resolution:} We sample both the ERA5 and HRRR data at 1-hour intervals. That is, we create a paired dataset at the same valid time $t_i$ of HRRR and ERA5 snapshots ($M_{t_i}, S_{t_i}$) with hourly time resolution. 

\textbf{Spatial Resolution:} The spatial resolution of HRRR and StormCast is 3km. We train StormCast on a spatial region over the Central US with spatial extent $1536$km $\times$ $1920$km Our vertical grid spacing is as fine as 125m in the boundary layer and as coarse as 500m in the free troposphere. The horizontal resolution of the ERA5 conditioning data is approximately 28km. 

\textbf{Data volume and train / test split:} Approximately 3.5 years (30,660 independent samples) of training data are used, from July 2018 through December 2021. Data for the year 2022 are held out for validation. Additional data during spring 2024 were used for testing in a forecast context, as follows.

\begin{table}[]
\begin{tabular}{l|l|l}
\multicolumn{3}{l}{ERA5}                                                                                                         \\
\hline
Parameter                                & Pressure levels (hPa)                                 &  Height levels (m)             \\
\hline
Zonal Wind (u)                           & 1000, 850, 500, 250                                   & 10                            \\
Meridional Wind (v)                      & 1000, 850, 500, 250                                   & 10                            \\
Geopotential Height (z)                  & 1000, 850, 500, 250                                   & None                          \\
Temperature (t)                          & 1000, 850, 500, 250                                   & 2                          \\
Humidity (q)                             & 1000, 850, 500, 250                                   & None                          \\
Total Column of Water Vapour (tcwv)      & Integrated                                            & -                             \\
Mean Sea Level Pressure (mslp)           & surface                                               & -                             \\
Surface Pressure (sp)                    & surface                                               & -                             \\
\hline
\multicolumn{3}{l}{HRRR}                                                                                                         \\
\hline
Parameter                                & Hybrid model levels (index)                           & Height levels (m)             \\
\hline
Zonal Wind (u)                           & 1, 2, 3, 4, 5, 6, 7, 8, 9, 10, 11, 13, 15, 20, 25, 30 & 10                            \\
Meridional Wind (v)                      & 1, 2, 3, 4, 5, 6, 7, 8, 9, 10, 11, 13, 15, 20, 25, 30 & 10                            \\
Geopotential Height (z)                  & 1, 2, 3, 4, 5, 6, 7, 8, 9, 10, 11, 13, 15, 20, 25, 30 & None                             \\
Temperature (t)                          & 1, 2, 3, 4, 5, 6, 7, 8, 9, 10, 11, 13, 15, 20, 25, 30 & 2                             \\
Humidity (q)                             & 1, 2, 3, 4, 5, 6, 7, 8, 9, 10, 11, 13, 15, 20, 25, 30 & None                             \\
Pressure (p)                             & 1, 2, 3, 4, 5, 6, 7, 8, 9, 10, 11, 13, 15, 20 & None                             \\
Max. Composite Radar Reflectivity (refc) & -                                                     & Integrated                    \\
Mean Sea Level Pressure (mslp)           & -                                                     & Surface               \\
Orography                                & -                                               & Surface \\
Land/Water Mask                          & -                                               & Surface \\
\end{tabular}
\caption{Parameters from the ERA5 and HRRR dataset that are used for training the StormCast model}
\label{tab:parameters}
\end{table}

\subsection{Forecast experiment design}

\textbf{Evaluation period:} A total of 135 StormCast forecasts are launched for the date range May 8 -- June 15 2024, four times daily at 0Z, 6Z, 12Z and 18Z. The beginning of this time interval was constrained by the onset of our own collection of synoptic conditioning data \( S_t \) from the real time NOAA Operational Model Archive and Distribution System~\cite{NOAANomads}, where it is cached at our desired hourly temporal frequency.

\textbf{Initialization and conditioning:} Each forecast used HRRR analysis states to initialize \( M_t\) to permit the HRRR data assimilation scheme to constrain the initial mesoscale state. Synoptic conditioning was provided via hourly output from a single deterministic $0.25^\circ$-resolution NCEP GFS global forecasts launched at common initial times. 

\textbf{Ensemble design:} At each initialization we generate a 5-member ensemble forecast using StormCast. At each 1-hour forecast step, the diffusion stage samples a forecast state starting from Gaussian random noise with variation in this sampling generating an ensemble of forecasts. The ensembles are propagated forward autoregressively at each timestep. Generating many more ensemble members is computationally cheap. However, probabilistic ensemble calibration requires further investigation. We expect to explore this aspect in future work.

\textbf{Forecast validation:} We compute forecast skill metrics over a small representative set of atmospheric variables -- composite reflectivity, surface zonal and meridional wind at 10m, surface temperature at 2m, as well as specific humidity, temperature, and zonal wind at hybrid levels 5 and 10 (approximately 700m and 2.5 km altitude). For all variables except composite reflectivity, we treat the HRRR analysis at the verification time as the ground truth. 

\textbf{Forecast validation (radar):} For composite reflectivity, we consider the Multi-Radar Multi Sensor~\cite{zhang2016multi} observation at the verification time to be the ground truth. Recognizing that precipitation scoring requires nuance, we validate radar using the Fractions Skill Score (FSS)~\cite{roberts2008scale, ebert2021cira}. The FSS compares the number of grid cells with precipitation exceeding a given threshold in the forecast and in the verification through a neighborhood of fixed size surrounding every point in the domain.  The FSS ranges between 0 and 1, with a perfect score of 1 meaning that the fraction of cells with precipitation above the threshold matched the verification through every neighborhood.  This approach avoids the double penalty that would be incurred using a score, such as RMSE, if a correct precipitation intensity is slightly misplaced relative to the verification.  

\textbf{Forecast validation (radar, ensemble): }To compare our StormCast ensemble forecast of radar reflectivity with the single deterministic HRRR forecast, we compute the Probability Matched Mean (PMM)~\cite{ebert2001ability, clark2017generation} of the FSS. The PMM has been shown in prior studies~\cite{schwartz2014characterizing} to be a more skilful estimate of a precipitation forecast from a CAM than individual ensemble members of the CAM as well as the plain ensemble average. 

\textbf{Intercomparing ensemble and determinstic results: } Unfortunately, corresponding ensemble forecasts from numerical models such as HRRR are computationally expensive and were not available to us in the operational archive. Since this means we do not have a fair physics-based CAM ensemble baseline to compare against StormCast ensemble forecasts, the appropriate interpretation of our comparisons of StormCast to HRRR is of the potential advantages of an ensemble generative machine learning approach. Our evaluation period had a small overlap of 15 days with the NOAA Hazardous Weather Testbed (HWT) Spring Forecasting Experiment (SFE)~\footnote{https://hwt.nssl.noaa.gov/sfe/2024/}. We obtained a small set of 5-member ensemble forecasts from the NSF-NCAR MPAS ensemble forecasts~\cite{NCAR_HWT2024, schwartz2024evaluation} run as a part of this experiment. The MPAS ensembles were not initialized with a radar-assimilating analysis like the HRRR and StormCast forecasts were, hence the forecasts are not directly comparable. Even so, we include a cursory comparison in Appendix~\ref{app:mpas} with the intention of expanding this analysis in future work. 

\section{Results}

We begin with a case study of forecasting a convective event  before discussing StormCast's statistical forecast skill, 
then analyze the multivariate relationships underlying its generated convection.

\subsection{Case Study}
\begin{figure}
    \centering
    \includegraphics[width=0.95\textwidth]{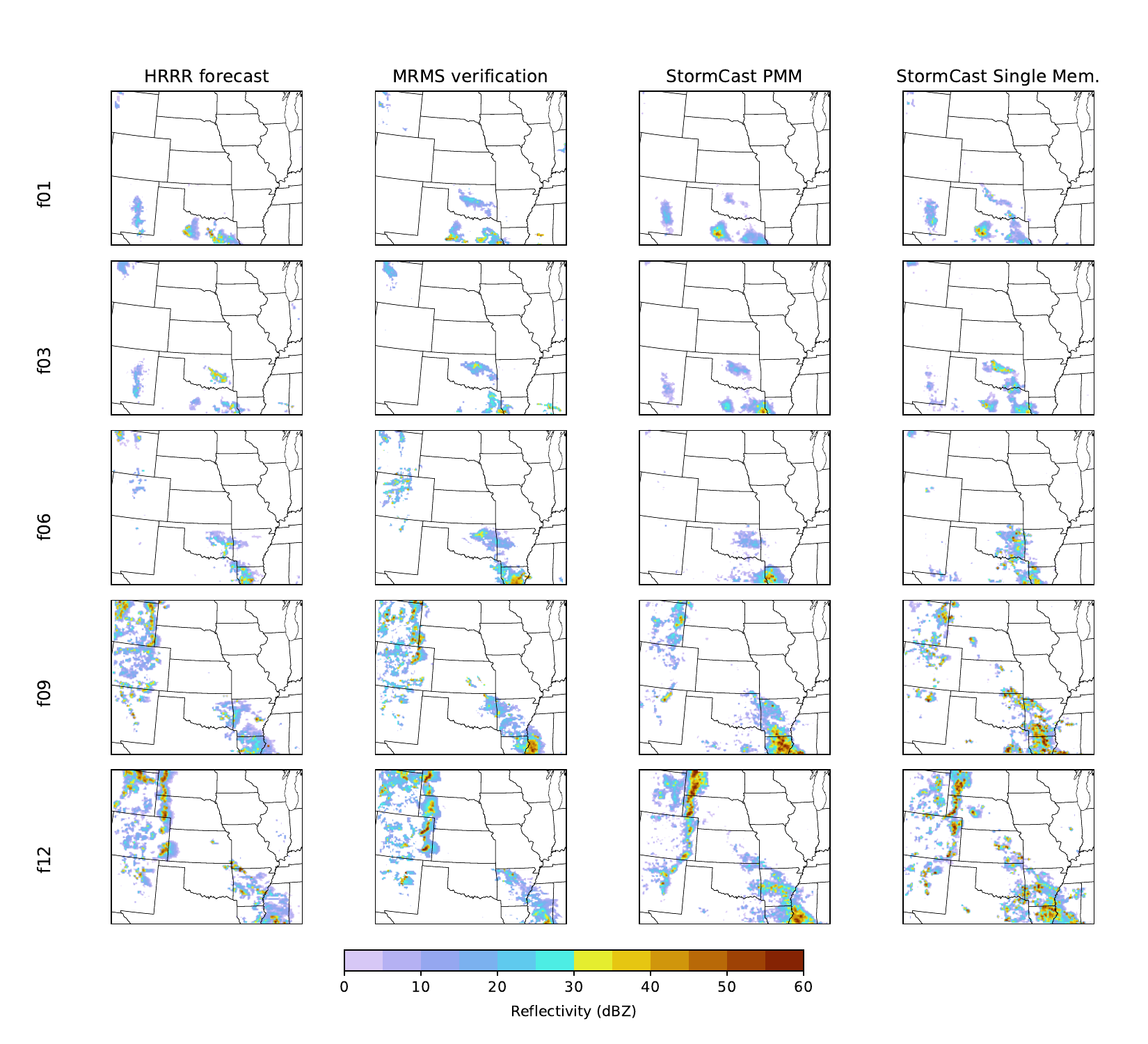}
    \caption{An example forecast of composite radar reflectivity generated by StormCast and HRRR compared with observed verification data from the Multi-Radar Multi-Sensor (MRMS) network. From left to right, the columns show the HRRR forecast, the corresponding MRMS observation, the Probability Matched Mean (PMM) of a five-member StormCast ensemble forecast and one of the StormCast ensemble members. The HRRR and StormCast forecasts were initialized at $2024$-$05$-$29$ $12:00$ UTC. The StormCast forecasts were initialized using the HRRR analysis at the initialization time. The rows from top to bottom show the forecast at progressively longer lead times (1 hour, 3 hours, 6 hours and 12 hours) along with the corresponding MRMS observation at the appropriate time.  }
    \label{fig:example_forecast}
\end{figure}

Figure~\ref{fig:example_forecast} shows that StormCast forecasts look qualitatively realistic, by comparing spatiotemporal details from an ensemble radar reflectivity forecast against a corresponding HRRR forecast and observed ground truth from the Multi-Radar-Multi-Sensor (MRMS)~\cite{zhang2016multi} network. The forecast begins at 12Z, early morning local time. The MRMS observations reveal two mature convective systems in the south of the domain, initially over central Oklahoma and northeast Texas, which propagate to the east, traveling about 600 km over the course of twelve hours while isolated afternoon convection builds over the western terrain of the Rocky Mountains, as is typical this time of year. Like the HRRR, StormCast successfully initializes, maintains and forecasts the propagation of these clusters, with some variation in the intensification of the southernmost system. Beyond the predictability horizon of the mesoscale, i.e. 6-12 hours into the forecast, both models generate variable internal representations of locally forced afternoon mountain convection. 
Commonalities across the ensemble members are captured in the StormCast PMM, which happens to validate better than HRRR for the southernmost system's observed intensification. Additional analogous case studies of forecasts are provided for a few more initialization dates in Appendix~\ref{app:case_studies_supp}\\

\subsection{Forecast skill}\label{sec:skill}

\begin{figure}[h]
    \centering
    \includegraphics[width=\textwidth]{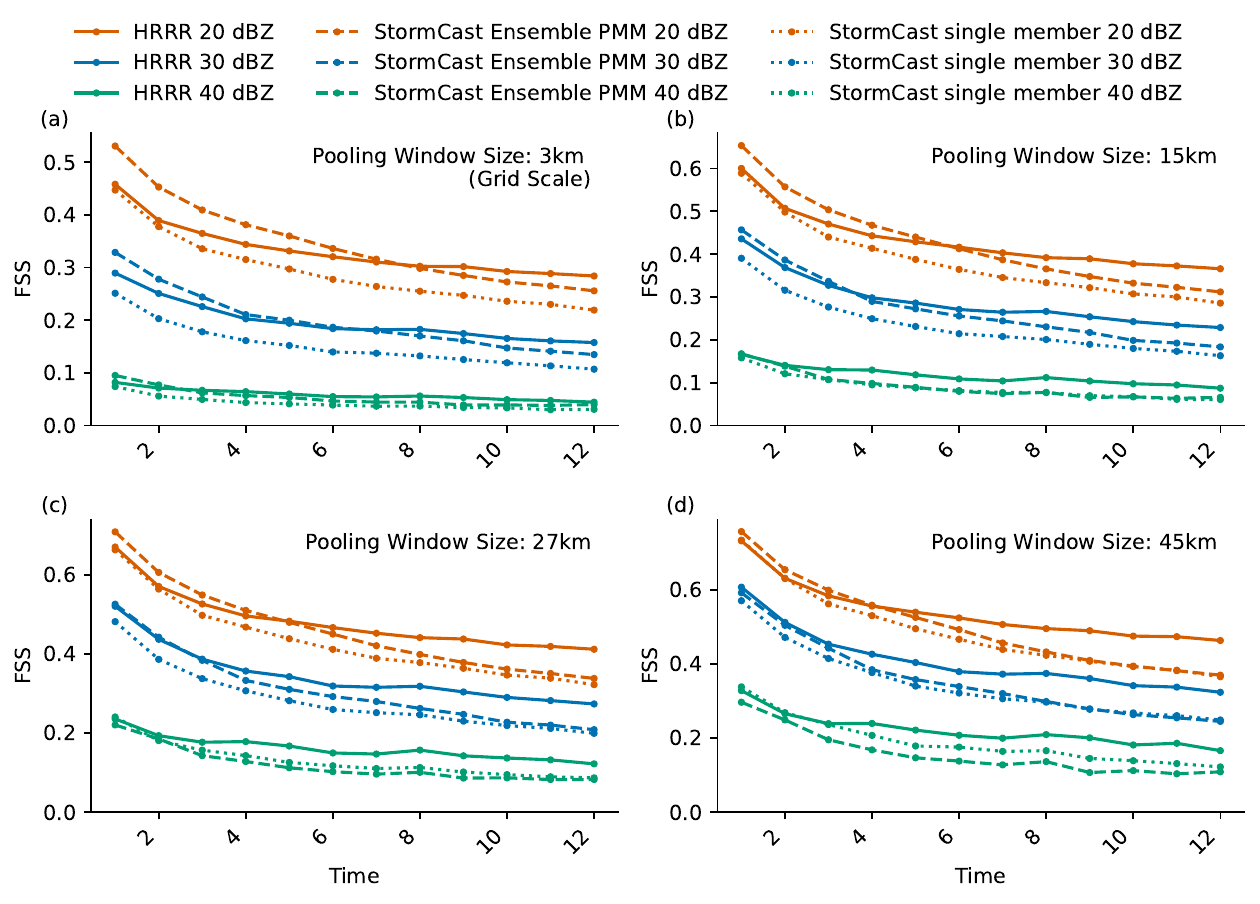}
    \caption{The Fractions Skill Score (FSS) of StormCast forecasts is compared with the corresponding FSS of HRRR forecasts. We compute the FSS at a few different pooling window sizes to illustrate the forecast skill at various spatial scales. }
    \label{fig:fss}
\end{figure}

 For forecasting precipitation, Fig.~\ref{fig:fss} shows that both the ensemble PMM and individual forecasts from StormCast have competitive FSS compared to HRRR in the first 12 hours. The FSS of composite reflectivity forecasts are evaluated at thresholds of 20dBZ (light rain), 30dBZ (light-to-moderate rain) and 40dBZ (moderate rain). We compare the single member forecasts from StormCast (dotted lines) as well as the PMM of a five-member ensemble from the StormCast model (dashed lines) against the HRRR single member forecast baseline (solid lines). 
 
 For the most predictable light rainfall category, the PMM of StormCast's ensemble FSS forecasts are more skillful than the HRRR out to lead times of 3 (5) hours for pooling window sizes of 45 (15) km (Fig.~\ref{fig:fss} c-d; orange dashed vs. solid); these larger spatial scales are also where intrinsic predictability is highest (i.e. FSS is systematically increased in magnitude). 
 StormCast's determinstic forecasts are also competitive with HRRR at 2 hour lead time for all pooling window sizes ( Fig.~\ref{fig:fss} a-d; orange dotted vs. solid), for this rainfall category. 
 
 For light-to-moderate rainfall, useful predictability (FSS larger than 0.4) exists only at 15-45 km pooling sizes on 1-3 hour lead times. Here, StormCast's ensemble skill appears to be competitive but not more skillfull than the HRRR(Fig.~\ref{fig:fss} c-d; blue dashed vs. solid) despite the fact that its individual predictions are systematically less skillful (Fig.~\ref{fig:fss}; blue dotted vs. solid). 
 
 For higher rain rates little useful predictability exists in either model, i.e. $FSS < 0.4$ at all lead times (Fig.~\ref{fig:fss} a-d; green lines). 
 
 So far, our analysis of StormCast's skill in predicting an especially important metric -- the FSS of radar reflectivity -- suggests that it is competitive with the HRRR, and has the capacity to become more skillful on lead times less than 6 hours, and that this is especially due to benefits of ensembling. 
 
\begin{figure}[h]
    \centering
    \includegraphics[width=\textwidth]{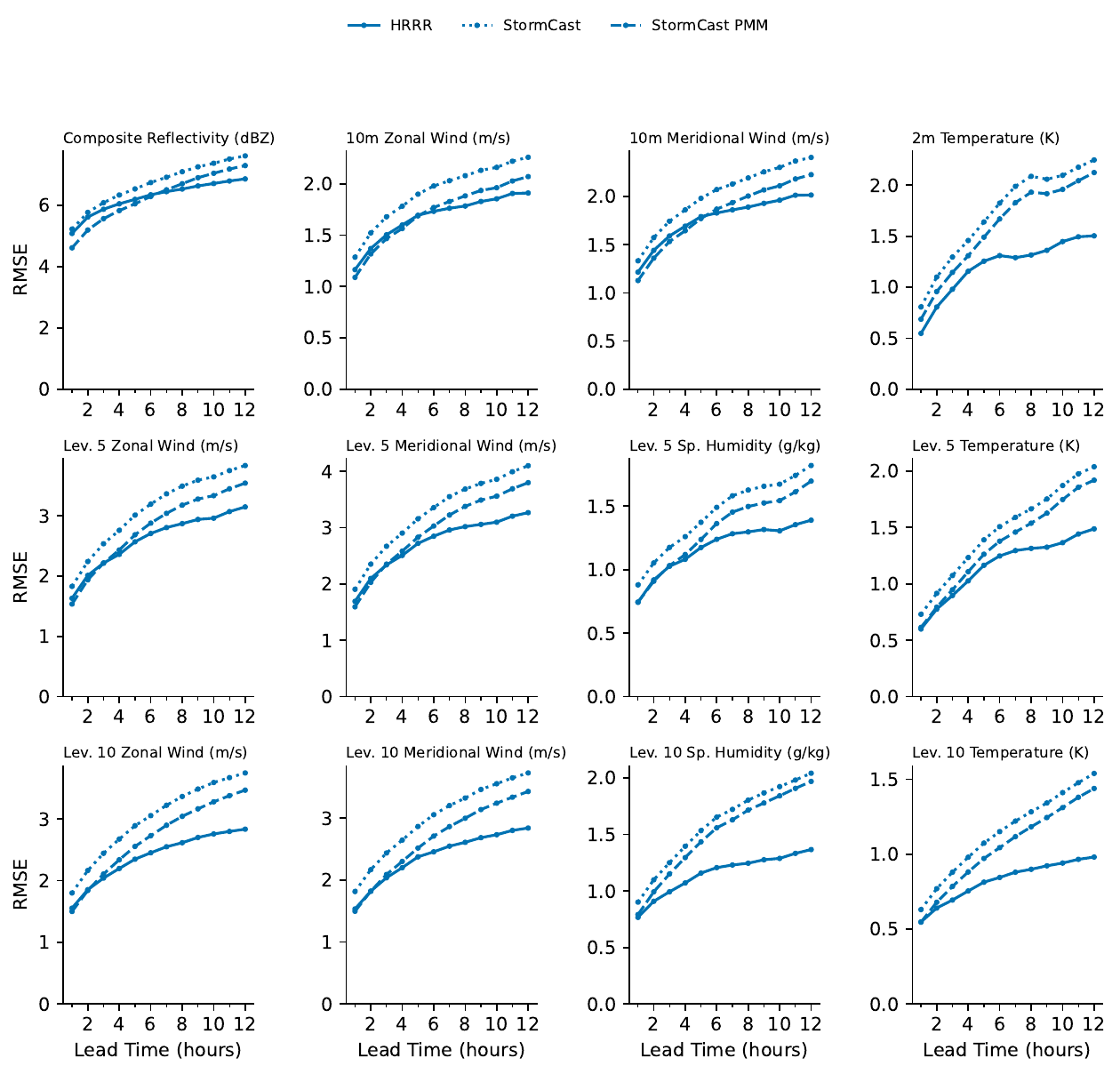}
    \caption{Forecast skill of a few selected variables predicted by the StormCast model compared with the HRRR model. Forecast skill is measured using the Root Mean Square Error between forecasts and the verification data for Composite Reflectivity, 10m wind velocity components, 2m temperature as well as the winds, temperature and specific humidity at HRRR native levels 5 and 10. The verification data for all variables except the composite reflectivity is the HRRR analysis at the verification time. For composite reflectivity, the verification data is the corresponding observed reflectivity from the MRMS sensor network at the verification time. The forecast RMSE scores are averaged over 130 forecasts from May 10 2024 to June 15 2024 with forecasts generated four times daily at 00, 06, 12, 18 UTC. We show the skill of the single member HRRR forecast (solid lines), a single member StormCast forecast (dotted lines), and the Probability Matched Mean (PMM) of a 5 member ensemble from StormCast (dashed lines).}
    \label{fig:rmse}
\end{figure}

 This finding is generalized across additional field variables in Fig. \ref{fig:rmse} by examining RMSE relative to the HRRR analysis. In general, the StormCast determinstic forecasts produce systematically higher errors and error growth rates than HRRR. However, on lead times less than 4-6 hours, the StormCast ensemble PMM is competitive with and capable of outperforming the HRRR. Of the variables analyzed, the StormCast PMM has \textit{less} error than the HRRR baseline for radar reflectivity (1-5 hour lead; consistent with FSS), and horizontal wind components near the surface (1-4 hour lead) as well as winds at two separate altitudes within the boundary layer (1-2 hour lead). Some issues  with other predicted variables are also uncovered: StormCast PMM predictions are systematically less skilful than HRRR for-near surface temperature and for specific humidity and temperature on the 10th model level, which is near 2.5 km altitude. At lower altitudes, on the 5th model level near 700 m altitude, errors in specific humidities and temperatures are comparable between the StormCast PMM and the HRRR baseline for short lead times.\\

 Overall our assessment is of encouragingly competitive precipitation forecast skill especially considering the crude ensemble initialization employed.

\subsection{Power Spectra and Probability Distributions}\label{sec:spectra}

We next investigate the quality of simulated power spectra and probability distributions during roll-out. For this task, we perform 14 separate 12-hour simulations initialized on a set of arbitrary dates during 2022, spanning a range of synoptic conditions. We use ERA5 reanalysis as conditioning for this and all subsequent analysis. We examine three single level variables and three variables at model level 10, which is around 1.5km height above the ground. 

\begin{figure}
   \centering
       \hspace{0mm}\includegraphics[width=0.9\textwidth,clip]{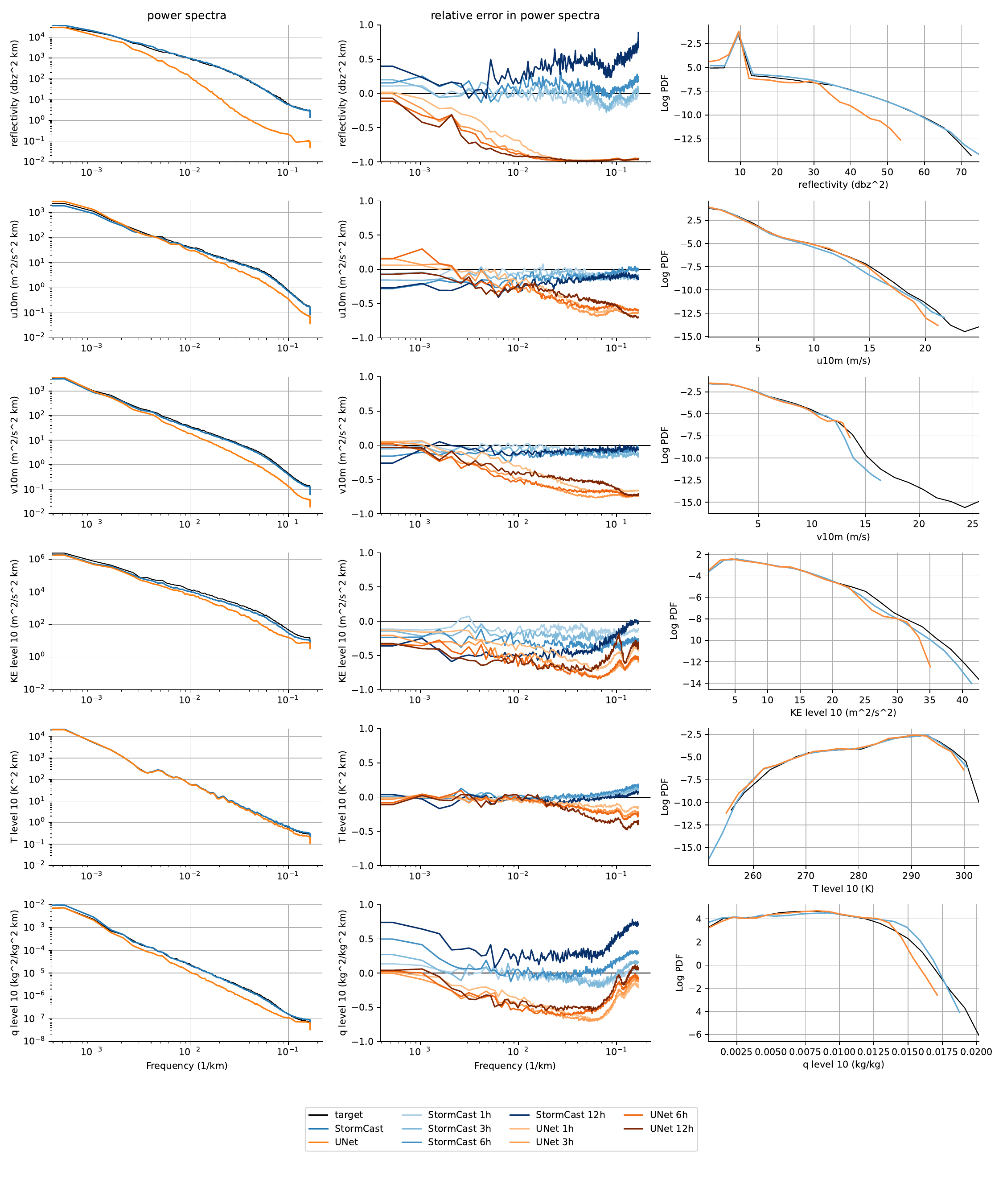}  
\caption{Power Spectra of selected variables. Left column shows the power spectra at lead time 3 hours comparing the diffusion model (blue) and the regression model (orange) with the target. Each row in the right column shows the relative difference (nondimensional) from the target (model/target-1) corresponding with the variable in the left column, for a selection of lead times. The top row includes also the spectra from the HRRR forecast for the corresponding times.}  \label{fig:spectra}
\end{figure}

The results show StormCast produces overall realistic spectra and probability distributions at a lead time of three hours (Fig. \ref{fig:spectra}, left). The diffusion model corrects for the expected blurring properties of the regression model, successfully injecting compensatory small-scale variance. Relative errors in power spectral density for the most skilfully predicted variables, radar reflectivity and low-level wind components, are less than 20\% in magnitude and exhibit encouragingly little scale-selectivity. 

\subsection{Qualitative assessment of convective dynamics}\label{sec:case_studies}

To further assess the extent to which the model accurately represents the physics of atmospheric convection, we perform a detailed multivariate evaluation of a few individual forecasts. This serves as a meteorological counterpart to assessing the trustworthiness and explainability of machine learning models. We examine the model's ability to produce vertically coherent and physically consistent convective motion, focusing on events where convective-scale dynamics should be evident. 

A challenge in examining convective-scale motions in the StormCast model arises from its 1-hour time resolution, which is too coarse to observe the temporal evolution of convection (e.g., through lagged correlation analysis or videos of individual updraft evolution). Consequently, our focus is on identifying the coexistence of characteristic convective features across different channels and vertical levels within the same time frame. This approach allows us to evaluate the model's ability to capture the spatial structure and inter-variable relationships typical of convective processes, despite the temporal resolution limitations.

\subsubsection{Multivariate Morphology of Precipitating Updrafts}\label{sec:updrafts}

Moist updrafts form when air in the boundary layer becomes warm and humid enough for rising pockets of air, called thermals, to continue ascending on their own~\cite{schlesinger1973numerical,schumann1991plume,barnes1995updraft}. As these thermals rise, water vapor within them condenses, releasing latent heat. This heat increases the air's buoyancy, propelling it further upward~\cite{kessler1974model,emanuel1997overview,park2017role}. The strength of these updrafts is directly related to the degree of buoyancy, which is influenced by temperature differences between the rising air and its surroundings~\cite{kessler1974model}. 
When these energetic thermals are strong enough to rise above the atmospheric boundary layer, they set the stage for precipitation. This occurs as cloud droplets form and grow within the rising air, a process that can be observed through radar reflectivity measurements~\cite{sassen1987ice,knight1993first,fridlind2015high}.

\begin{figure}
    \centering
        \hspace{0mm}\includegraphics[width=\textwidth,clip]{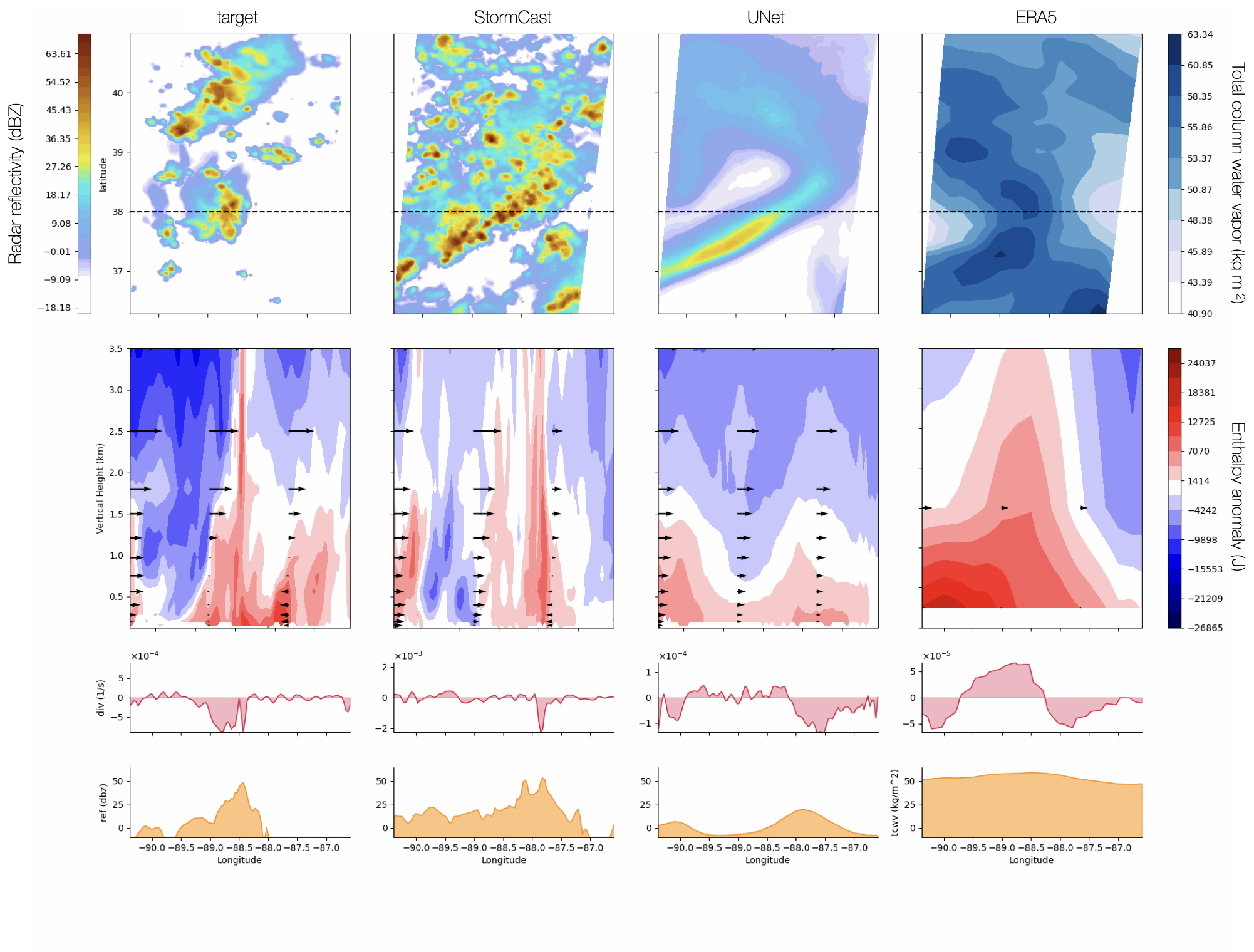}  
\caption{Examining updrafts in a free summertime convection event in Illinois on 2022-07-09 00:00:00 UTC at a lead time of 6h into the simulation. Left to right: target, StormCast, regression and ERA5. Top row: planar view of cloud related variable: radar reflectivity (dBZ) for the first three rows and total column water vapor (unit) for the last row. The dashed black line at 38N indicates the zonal section along which the lower figures are plotted. Second row: zonal-vertical slice showing contours the anomalies from a mean profile of the virtual moist enthalpy overlaid with zonal wind vectors. Third row: a vertical average of the horizontal divergence in the boundary layer (first 7 layers corresponding to first km) as a function of longitude. Fourth row: section of the radar reflectivity. For graphical simplicity we do not include the colorbars for the reflectively and total column water vapor panels at the top row, but their range of values could be inferred from their sections in the lowest row.}
\label{fig:free_conv_sec}
\end{figure}

Figure \ref{fig:free_conv_sec} examines such plumes from several vantage points, at 6h lead time during a time period when environmental conditions were prone to convection with weak synoptic forcing on 2022-07-09 00:00:00 UTC, focusing on a small  $4^\circ \times 4^\circ$ region over the state of Illinois. The top row depicts a planar view of radar reflectivity, showing the cluster of disorganized convection that occurred in observations, alongside a StormCast ensemble member that predicted a linear form of organized convection oriented from northwest to southeast in a similar location \footnote{The purpose is not to assess any match between km-scale details of the observations and predictions -- none should be expected for these spatial scales at a 6h lead time -- but rather to anchor an analysis of StormCast's own internally generated convective motions.}. A dashed line indicates a convenient latitude where convection occurred in both observations and StormCast, which defines the vertical-zonal section examined in all panels below. 

The second row of Fig. \ref{fig:free_conv_sec} reveals vertical structure along this dashed line, with shaded contours representing the anomaly of virtual moist enthalpy in the height-longitude plane, with respect to a reference profile. Moist enthalpy is a measure of the combined heat and vapor that provides buoyancy to deep convective updrafts. It is a useful quasi-conserved quantity under adiabatic vertical motions and liquid-vapor phase changes of water at saturation.

To help identify the relative location of updrafts, black vectors represent zonal wind, and the third row of Fig. \ref{fig:free_conv_sec} displays the net boundary layer divergence. This is a convenient proxy for vertical motion, which was not directly modeled in StormCast. By mass conservation, net BL divergence (convergence) corresponds to downward (upward) motion at the top of the boundary layer since air cannot flow through Earth's surface. Net BL divergence is estimated as the average across the model's seven lowermost vertical levels. As with enthalpy, the horizontal mean is removed to isolate smaller-scale features from the large scale subsidence and synoptic environment. 

Finally, to locate the regions of precipitation, the fourth row shows a zonal transect of radar reflectivity. 

Our physical expectations from observations and cloud-resolving simulations are that precipitation should spatially co-locate with intense updrafts and positive enthalpy anomalies protruding above the boundary layer. This is true of the target data (left column), where peak radar reflectivity at -88.5 deg E coincides with a prominent implicit updraft (i.e. negative net BL divergence anomaly; low-level convergence) and a positive enthalpy anomaly that extends from the boundary layer up to 3km altitude, at the same longitude.

Our expectations are also confirmed in StormCast output (second column from left). StormCast predicts peak radar anomalies to occur one degree farther east, near -88 deg E. A negative net BL divergence anomaly and a notable postitive enthalpy anomaly extending beyond 2 km in altitude occur at the same longitude. Two other meteorological conditions are also examined in Appendix \ref{app:cases}. 

The rightmost columns of Fig. \ref{fig:free_conv_sec} attest to the importance of the generative diffusion component of StormCast's architecture in producing these dynamics. While the regression model helps condition the overall location of peak convection based on conditioning from ERA5, only the generative model produces multi-variate km-scale updraft anomalies of realistic magnitude -- less than $-10^-3$ s$^{-1}$ for net divergence, more than 40 dBZ for radar reflectivity, and positive enthalpy greater than 3 kJ extending up to 3 km.

\subsubsection{Cold Pools}\label{sec:cold_pools}

Atmospheric cold pools formed in the vicinity of convection are dense masses of air cooler than the surrounding environment which descend from cloud bases and spread out along the ground~\cite{thorpe1982}. Cold pools are created when precipitation evaporates as it falls through the drier air underneath the convective storm~\cite{houze1981convection}. This evaporation cools the surrounding air, making it denser than the environment, leading to the formation of a density current near the surface. This cooler air spreads outward horizontally, forming a roughly circular/oval shape. Typical characteristics of cold pools are that their in-pool temperature is typically $1.5^\circ C$-$10 ^\circ C$ cooler than the environmental air~\cite{fujita1959precipitation, droegemeier1985three, tompkins2001organization}. In addition, they are accompanied by strong, gusty winds at the leading edge, forming a gust-front signature near the surface~\cite{droegemeier1985three}. Their average diameter can range from a few kilometers (from small convective cores) to over 100 km (related to MCS or squall line thunderstorms)~\cite{feng2015mechanisms}. Cold pools play an integral role in controlling thunderstorm dynamics and can trigger new convection by lifting warm, moist air at their boundaries~\cite{fujita1959precipitation}. Although they are important in controlling the convective dynamics in an unstable environment, these mesoscale features are not well represented in current weather models~\cite{grant2024}. 

\begin{figure}
    \centering
        \hspace{0mm}\includegraphics[width=\textwidth,clip]{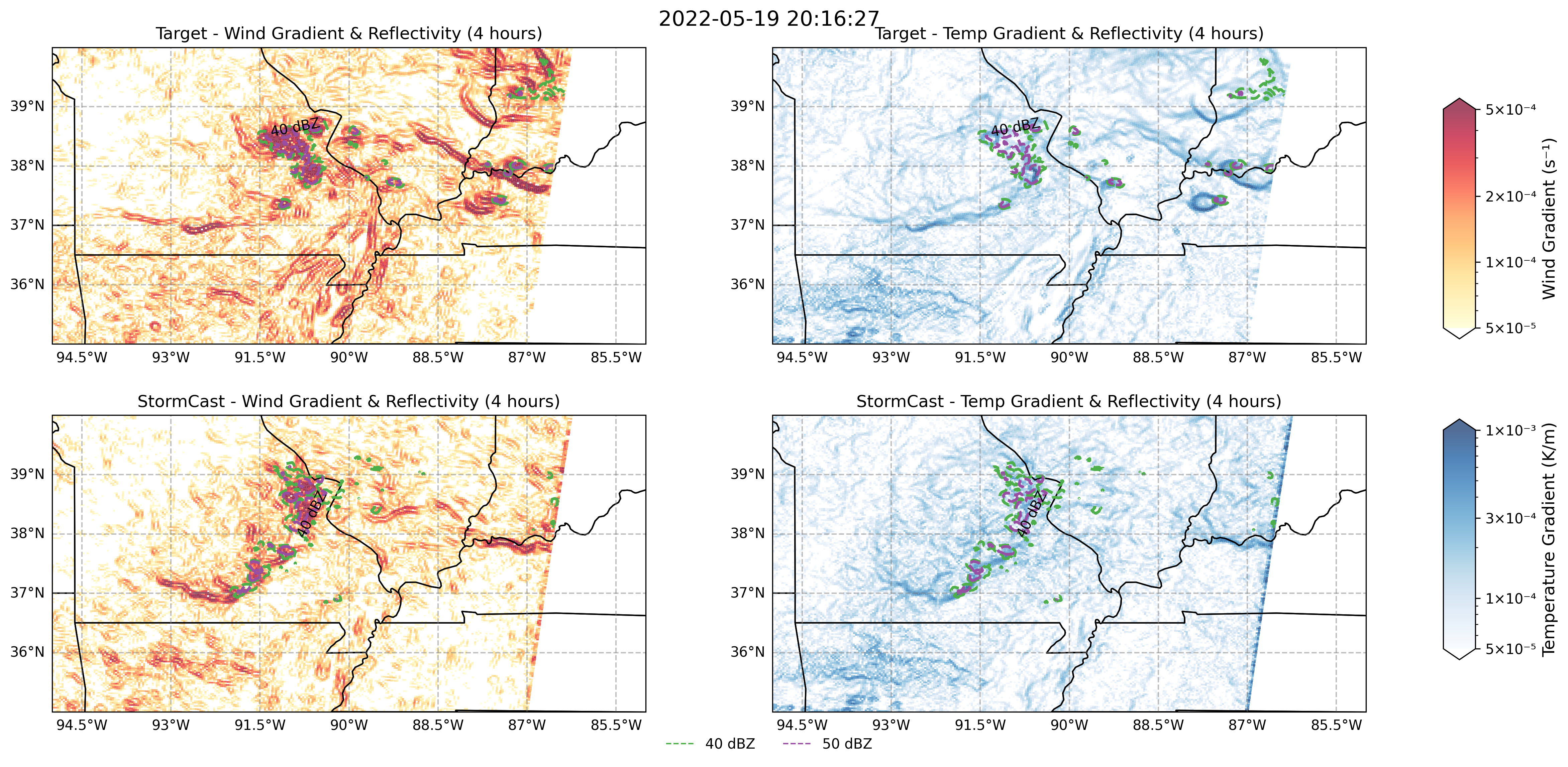}  
\caption{Representative fine-scale features of the horizontal gradients of 10m horizontal wind (left column) and 125m air temperature gradient (right column) for target (top rows) and StormCast 4-hour forecasts (bottom rows), suggestive of cold pool related gust fronts. This case was initialized on 2022-05-19 17:30:00 UTC. Composite radar reflectivity contours at 40 and 50 dBZ are overlain in green and purple.}  \label{fig:wind_temp_grad}
\end{figure}

The presence of cold pools in StormCast simulations is a useful test of its ability to learn an implicit representation of the relationship between precipitation production and evaporative cooling that is the source of negative buoyancy for cold pools. The co-development of thermal fronts and surface wind anomalies by cold pools is another test of multi-variate physical consistency in the model. To evaluate this, we identify cold pool related gust front signatures in the 10m near-surface wind fields (u10m and v10m) and air temperature at 125 m above the Earth's surface. We calculate horizontal wind and temperature gradients~\cite{garg2020identifying} using finite differences, followed by edge enhancement using Sobel filter~\cite{gao2010improved}. Figure \ref{fig:wind_temp_grad} displays composite reflectivity contours at 40 and 50 dBZ with scalar wind and temperature gradient fields for both the target (top rows) and StormCast (bottom rows). These fields are depicted at 4 hours in the simulation to determine if StormCast has learned the cold pool representation or is merely deriving it from initial conditions.

Cold pools are known to have a distinct signature in temperature and wind fields~\cite{tompkins2001organization,garg2020identifying} as the gust fronts resulting from rain evaporation exhibit stronger winds and lower temperatures compared to the surrounding air mass. Figure \ref{fig:wind_temp_grad} depicts distinct arc-like features of maximized gradient in both fields surrounding convection in the target HRRR data between Illinois and Missouri. Likewise, StormCast 4-hour forecasts contain arc-like structures surrounding portions of its most intense generated radar reflectivity feature over Missouri. Another gust front signature between Illinois and Indiana is clearly visible in both target and StormCast gradient fields. As in the target data, some co-location in both the temperature and horizontal wind gradient maxima are apparent in StormCast. Appendix \ref{app:cold_pool} contains additional analysis of objectively identified gust fronts. 

\section{Discussion}
The twin challenges of emulating realistic km-scale 3D atmospheric dynamics, and preserving fine-scale stochastic details under the constraint of relatively coarse temporal resolution (hourly) data required two innovations: (1) developing a generative ML emulator; (2) predicting a dense atmospheric state with dozens of vertical levels. StormCast, the first-of-its-kind generative ML emulator of a km-scale convection-allowing model shows that generative ML models have the potential to greatly improve and augment mesoscale meteorology in the same way that deterministic ML has improved meteorology at coarser resolutions. 

We are encouraged by the following key findings: \\
1. StormCast's forecasts are skilful and competitive with the operational HRRR model. \\
2. Unlike statistical nowcasting, StormCast achieves this while generating multiple realizations of the evolution of mesoscale systems, cold pools, and moist updrafts with realistically co-located low-level convergence, mid-tropospheric enthalpy and radar reflectivity anomalies. \\
3. StormCast's predicted mesoscale dynamics validate well on power spectra and probability distributions.\\
4. StormCast's small, 5-member ensemble can create an ensemble PMM with better skill than a single physics-based forecast, including by directly leveraging HRRR data assimilation through a consistent initialization.\\
5. Although trained with synoptic conditioning from ERA5, StormCast is robust to the shift to deterministic GFS forecast conditioning during test time.\\

Taken together, these results imply strong potential for generative autoregressive ML as a valid alternative to traditional km-scale numerical prediction -- enabling new applications in weather forecasting and climate downscaling alike.

We readily admit several limitations of our model worthy of future work. StormCast was trained on a relatively small amount of data from 3.5 years of HRRR analysis. Increasing the volume of training data could improve the forecast accuracy of future models. Similarly, increasing the domain size of training could expose the model to a more diverse set of severe weather phenomena to learn from. Reanalysis datasets such as CONUS404~\cite{rasmussen2023conus404} offer much longer historical records and could be leveraged for training to improve model performance. We inadvertently trained on forecast state data lagged one hour from the HRRR data assimilation cycle; training directly on data assimilated state data could produce complementary skill. Our evaluation period was limited to a short time interval spanning only a portion of spring of 2024; expanding skill assessment across more regions and phases of the seasonal cycle could uncover additional findings. Finally, ensuring consistency in the training and testing data for conditioning could improve results.

Preliminary evaluations indicate that StormCast ensembles are underdispersive (Appendix~\ref{app:calibration}). Better training and sampling strategies for diffusion models and incorporating sources of initial condition uncertainty could alleviate this issue. Further, information about the synoptic-scale forecast uncertainty could be incorporated into StormCast by driving the StormCast model with a global ensemble forecast such as the NCEP Global Ensemble Forecast System (GEFS) rather than driving StormCast with a single GFS forecast. Further improving ensemble calibration could unlock the path to very large operational ML ensemble forecasts at the km-scale that would otherwise incur prohibitive computational costs with current numerical models.

Further work is needed to evaluate the physical consistency and statistical skill of mesoscale ML prediction models such as StormCast. This requires a community effort in designing appropriate forecast baselines and metrics. The performance of ML models in accurately forecasting severe weather must be rigorously tested before such models can be used in an operational setting. The role of meteorologists and atmospheric scientists will be prominent in designing benchmarks for mesoscale weather forecasts, particularly in assessing forecast skill on severe weather. Leading Numerical Weather Forecast centers such as the European Center for Medium Range Weather Forecasting (ECMWF) and the US National Oceanic and Atmospheric Administration (NOAA) have played a role both as creators of state-of-the-art ML models~\cite{lang2024aifs} as well as the trustworthy arbiters of the forecast quality of novel experimental ML forecast paradigms emerging from academia, industry and operational centers. To achieve its full potential, new simulation datasets at various length and time scales will need to be created tailored to ML model training. We hope that the results presented in our work inspire the development of regional analysis, re-analysis and re-forecast datasets over different parts of the globe for training ML models such as StormCast.

Our current work focuses on model emulation alone. Generating a reliable weather forecast requires the additional critical step of assimilating observations to generate an estimate of the present state of the earth system, from which a forecast can be generated (c.f. Ref.~\cite{gustafsson2018survey}). We rely on the numerical HRRR model for state estimation and our model is therefore not an end-to-end ML-based forecast system. However, recent studies have proposed applications of generative diffusion models for data assimilation at a range of spatial scales, including at km-scale resolution~\cite{qu2024deep, huang2024diffda, manshausen2024generative}. Leveraging these advances could allow for an ML-based end-to-end forecast system which could greatly accelerate and augment operational time-critical ensemble generation at km scales and beyond.

\section{Acknowledgements}
We thank the National Oceanic and Atmospheric Administration (NOAA) and the European Center for Medium-Range Weather Forecasting (ECMWF) for sharing data and research openly. We thank Imme Ebert-Uphoff and researchers at the Colorado State University Cooperative Institute for Research in the Atmosphere (CIRA) for helpful comments and discussions. We thank Tom Hamill and The Weather Company (TWC) for constructive feedback. We thank Christopher Bretherton and the Allen Institute for Artificial Intelligence for helpful discussions. We thank scientists at the National Severe Storms Laboratory (NSSL) and the Storm Prediction Center (SPC) for helpful discussion on storm-scale weather. We thank Craig Schwartz and the NSF-NCAR HWT Spring forecast ensemble team for providing access to MPAS ensemble forecasts. This research used resources of the National Energy Research Scientific Computing Center (NERSC), a Department of Energy Office of Science User Facility using NERSC award ERCAP0028849.

\clearpage
\bibliographystyle{unsrt}

\clearpage
\appendix
\newcounter{AppendixCounter}
\renewcommand{\theAppendixCounter}{\Alph{AppendixCounter}} 
\renewcommand\thefigure{\Alph{AppendixCounter}\arabic{figure}}

\refstepcounter{AppendixCounter} 
\section*{Appendix \theAppendixCounter: Case Studies}\label{app:cases}
\setcounter{figure}{0}
We complement the analysis in Figure \ref{fig:free_conv_sec} with two additional figures from different meteorological scenarios. 
Figure \ref{fig:mcs_sec} illustrates the 6th hour of a simulated nocturnal Mesoscale Convective System (MCS) over Missouri. The convective system is discernible in both the planar view of radar reflectivity (top row) and the vertical-zonal slice of moist enthalpy (second row). 
It is encouraging that 6 hours into a forecast StormCast produces a plausible MCS reflectivity morphology including a core of intense convection embedded within a larger region of lower radar reflectivity reminiscent of a leading virga anvil. 
The eastward-moving cloud system is characterized by a positive enthalpy anomaly concentrated in the eastern section of its cloud area, contrasting sharply with a negative enthalpy anomaly in its wake, consistent with a trailing mesocale downdraft. Realistic generated multivariate updraft morphology is apparent at two longitudes of peak radar reflectivity (-89.5 deg E and -87 deg E) where plumes of enthalpy extending into the boundary layer co-locate with net BL convergence and peaks in radar reflectivity .
Figure \ref{fig:squall_line_sec} depicts the 6th hour of a simulated squall line over Texas, representing a clear case of a synoptically driven system. Again, the diffusion component of StormCast is found to produce spatially co-located km-scale anomalies in radar reflectivity, net BL convergence, and penetrative enthalpy plumes. 

These additional cases demonstrate the model's capability to capture diverse convective structures and highlight the varying degrees of improvement offered by the diffusion model across different meteorological conditions. The analysis underscores the model's strength in enhancing fine-scale features, particularly in radar reflectivity fields, while maintaining consistency with larger-scale structures present in the conditioning data.

\begin{figure}
    \centering
        \hspace{0mm}\includegraphics[width=0.8\textwidth,clip]{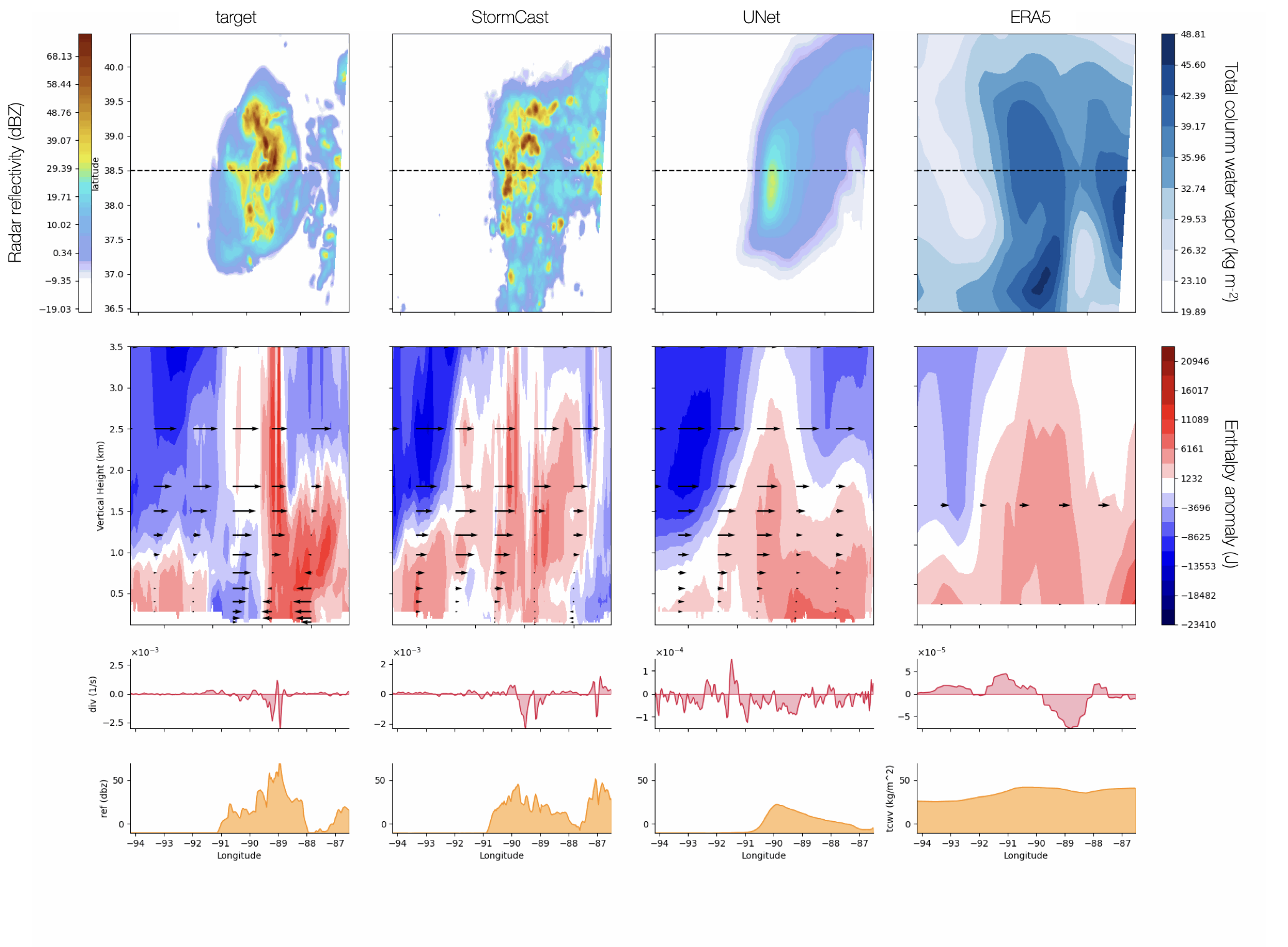}  
\caption{Examining a nocturnal MCS in Missouri on 2022-05-19 23:00 UTC at a lead time of 6h into the simulation. The figure format follows that of  \ref{fig:free_conv_sec}.}
\label{fig:mcs_sec}
\end{figure}

\begin{figure}
    \centering
        \hspace{0mm}\includegraphics[width=0.8\textwidth,clip]{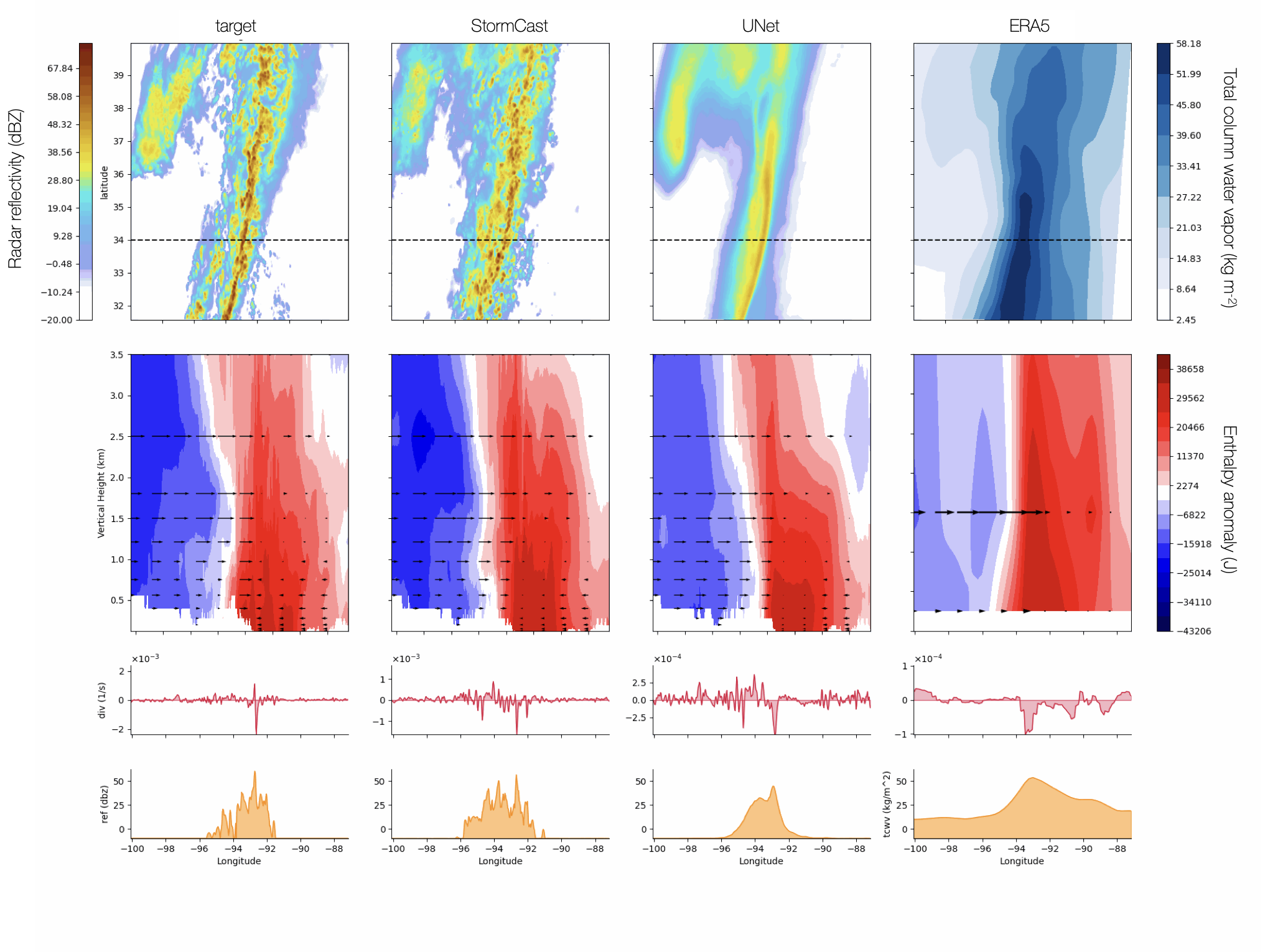}  
\caption{Examining a squall-line passing in Texas on 2022-11-05 03:00 UTC at a lead time of 6h into the simulation. The figure format follows that of \ref{fig:free_conv_sec}}
\label{fig:squall_line_sec}
\end{figure}

\clearpage

\refstepcounter{AppendixCounter} 
\section*{Appendix \theAppendixCounter: Cold Pool Identification}\label{app:cold_pool}
\setcounter{figure}{0}
We identify cold pool related gust front signatures in the 10m near-surface wind fields (u10m and v10m). The cold pool identification algorithm begins with calculating horizontal wind gradients using 10m surface winds~\cite{garg2020identifying} using finite differences, followed by edge enhancement using Sobel filter~\cite{gao2010improved}. We use two wind gradient thresholds here, first one at $1.85 \times 10^{-4}$ and second at $9.25 \times 10^{-5}$. These dual thresholds help us focus on mesoscale features without compromising on large scale synoptic circulations. A scale-space representation~\cite{lindeberg1994scale} is created using Gaussian filters at multiple scales, and differences between scales are computed to highlight sharp features. Scale-space representation is a technique used in image processing and computer vision to analyze images at multiple scales. It creates a family of derived images that represent the original image at different levels or blur or detail. This multi-scale representation allows for the detection and analysis of features that may be prominent at different scales within the image. The basic idea behind scale-space representation if to apply a series of smoothing operators to an image (typically Gaussian filters with increasing sigma). In this cold pool identification algorithm, this scale-space representation helped identify features by considering multiple scales, which made the edge detection more robust to noise and variations in feature size. The Canny edge detection algorithm~\cite{rong2014improved} is then applied to identify potential edges. These edges are refined using reflectivity-based masking, with morphological operations used to create buffered areas of interest. This process filters out small-scale noise and static non-convective features using connected component analysis, retaining only cold pool-related edges. 

\begin{figure}
    \centering
        \hspace{0mm}\includegraphics[width=\textwidth,clip]{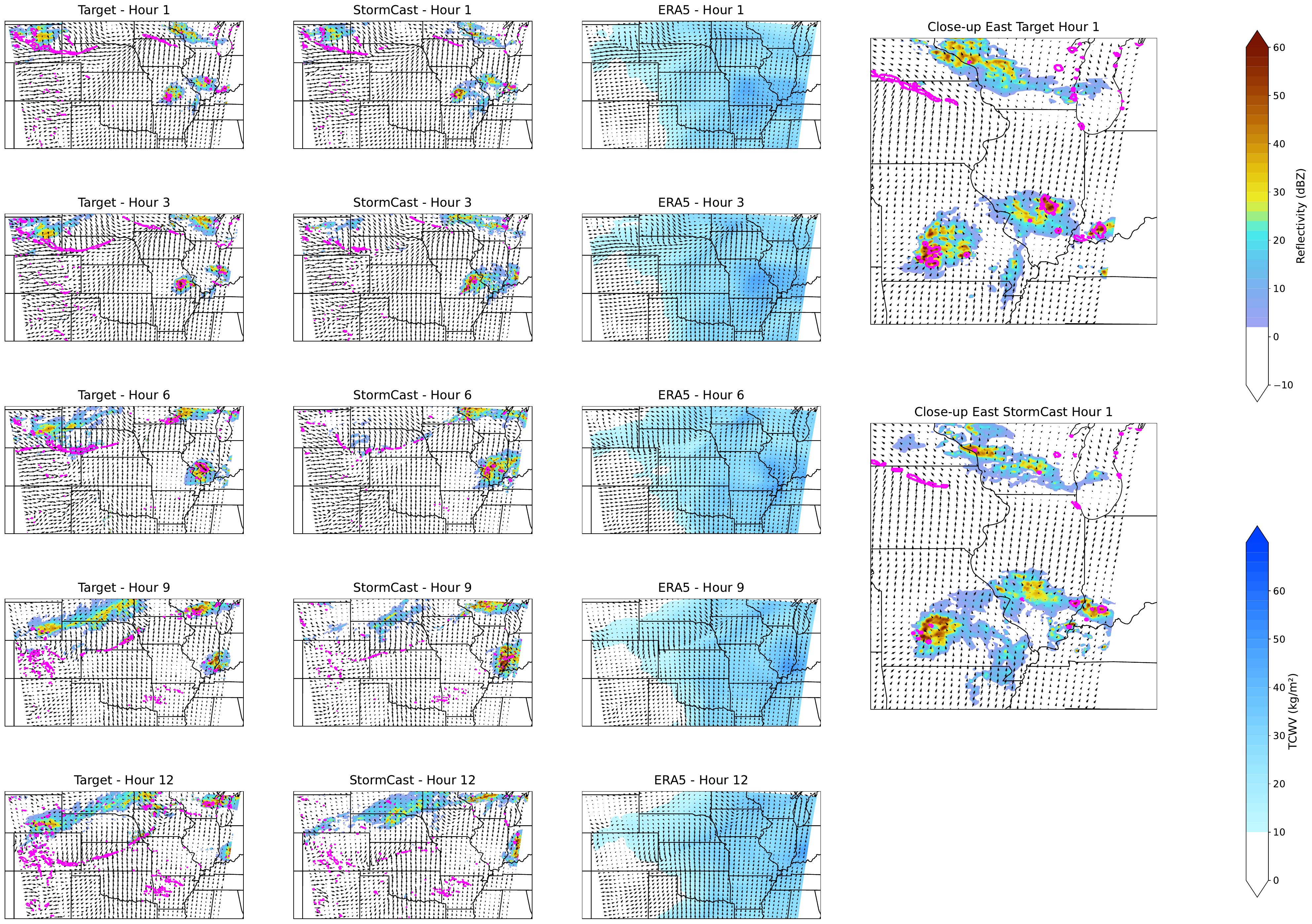}  
\caption{Examining cold pools during summertime convection over the domain initialized on 2022-05-19 17:30:00 UTC. These panels depict the evolution of convective systems showing composite radar reflectivity (dBZ) contours, 10m near-surface winds and cold pool-associated gust fronts (magenta contours) for the target and StormCast. The first column shows the evolution for the target data and the second column is for the StormCast predictions. The third column shows the ERA5 TCWV with 10m winds. The rightmost column shows the close up around the eastern boundary of our domain with images of target and StormCast fields at hour 1.}  \label{fig:free_conv_cold_pools}
\end{figure}

Figure \ref{fig:free_conv_cold_pools} shows composite reflectivity contours with 10m near-surface wind vectors and cold pool boundaries (magenta contours) at 3-hour intervals over the 12-hour forecasting period for target (HRRR) and StormCast. We also depict TCWV from ERA5 with their 10m winds in the third column. The motivation behind performing this analysis for ERA5 winds is to identify if the mesoscale cold pools are derived from the large-scale ERA5 fields or they are actually resolved by StormCast. At hour 1, target and StormCast depict a strong outflow in the northwest, between Nebraska and Wyoming, characterized by distinct wind shifts near the cold pool boundaries. This boundary is also visible in ERA5 winds but no gust front lines passing our gradient threshold identification algorithm occur due to their relative bluriness. An MCS between Missouri and Illinois shows some mesoscale cold pools related to this system. The close-up figure centered around these cold pools in the rightmost panel clarifies the strong outflow in the vicinity of this subregion of convection. Another cold pool boundary is visible between Minnesota and Iowa. As the simulation progresses, StormCast demonstrates proficiency in resolving these mesoscale structures, closely mirroring the target. However, it appears  that StormCast produces gust front signatures that are not as cohesive horizontally as the target data after the first hour of the simulation. 
\clearpage

\refstepcounter{AppendixCounter} 
\section*{Appendix \theAppendixCounter: Diffusion Model Configuration}\label{app:diffusion}
\setcounter{figure}{0}
Denoising diffusion models are designed to model a given distribution $p_{\mathrm{data}} (x)$ by successively denoising a series of noised versions of samples $x \sim p_{\rm{data}}$ from the data distribution. The noised versions are created by adding identically distributed Gaussian noise with standard deviation $\sigma_i$ to $x$, where increasing $\sigma_i$ values correspond to increasingly noised samples (asymptoting to pure white noise as $\sigma_i$ approaches $\sigma_{\rm{max}} \gg \sigma_{\rm{data}}$). During each training iteration, the model is given a noised sample from $\bf{x}_i$ and is trained to remove the added noise; during sampling it can then start with a sample of pure white noise and iteratively denoise it to obtain a sample from the training distribution.

In designing our diffusion model we closely follow the architecture, design and training protocol proposed in Ref.~ \cite{karras2022elucidating}, which carefully reformulates the training objective, noise distributions, and sampling procedures to improve training stability and sample quality. Our goal as outlined in Sec.~\ref{sec:prob_forecast} is to generate the conditional distribution $p(r_{t+1} \mid \mu_{t+1}, M_t, S_t)$ using a diffusion model. We wish to sample the residual $r_{t+1}$ conditioned on the deterministic mean forecast $\mu_{t+1}$ and the previous mesoscale and synoptic-scale states $M_t$, $S_t$. We denote the conditioning by
\begin{align}
    y \coloneq \left[ \mu_{t+1} ; M_t ; S_t \right],
\end{align}
where the $;$ indicates concatenation.

The diffusion training objective to be minimized can then be written as 
\begin{equation}
    \mathbb{E}_{\sigma, r, n} [ \lambda(\sigma) \| D_\phi (r + n ; y ; \sigma) - r   \|_2^2  ],
\end{equation}
where the expectation is taken over noise levels $\sigma$ sampled from the training noise distribution $p_{\mathrm{train}}$, samples $r$ from the target data distribution $p_{\mathrm{data}}$, and gaussian noise $n \sim \mathcal{N} (\mathbf{0}, \sigma^2 \mathbf{I})$. We select the DDPM++ U-Net architecture \cite{song2021scorebased} as the backbone for the denoiser. The network has 6 encoder and decoder layers with a base embedding dimension of 128. The training noise distribution $p_{\rm{train}}$ was chosen to be a log-normal distribution with mean $-1.2$ and standard deviation $1.2$. We train the network for a total of 450,000 training steps with batch size 64, corresponding to 29 million noised samples seen in training. The training pipeline completes in 120 hours on 64 NVIDIA H100 GPUs. At inference time, we initialize the random latent with Gaussian noise with a standard deviation $\sigma=800$ and use the second order solver proposed in Ref.~\cite{karras2022elucidating} for 18 denoising steps to generate a sample given an input condition. 

\clearpage

\refstepcounter{AppendixCounter} 
\section*{Appendix \theAppendixCounter: Ensemble Calibration}\label{app:calibration}
\setcounter{figure}{0}

\begin{figure}[h]
    \centering
    \includegraphics[width=\textwidth]{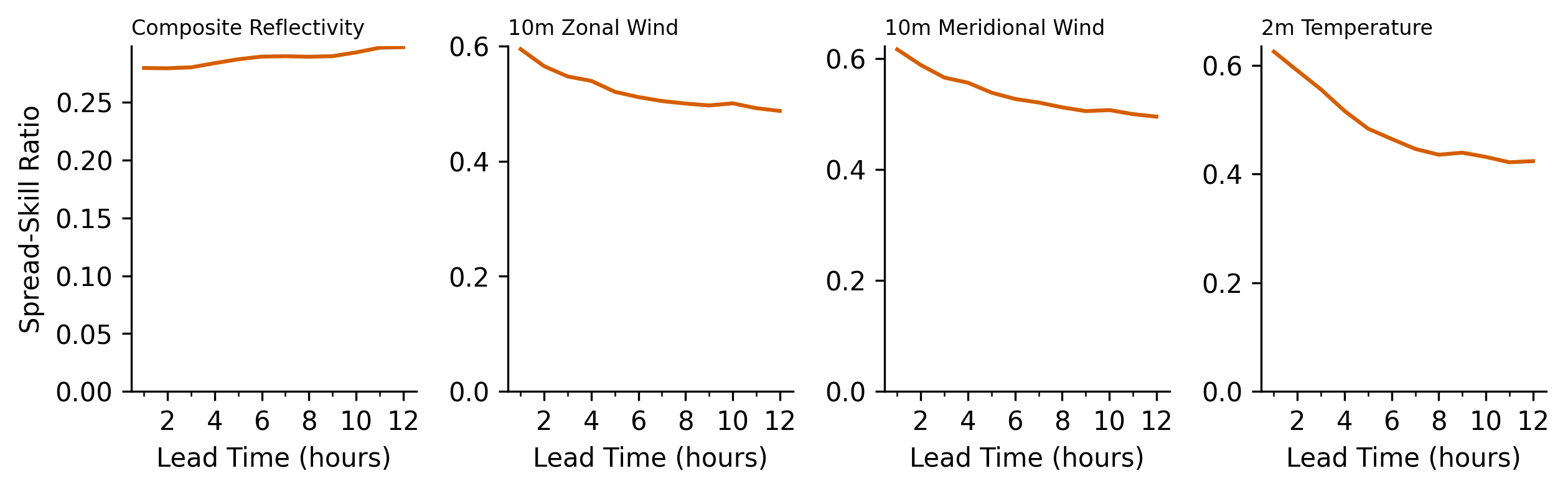}
    \caption{The spread-error ratio of a 5-member ensemble forecast made by StormCast indicating an underdispersive ensemble generated by StormCast. }
    \label{fig:spread-skill}
\end{figure}
We perform a preliminary analysis of the calibration of the small ensemble forecasts generated by StormCast. We compute the ratio of the enesemble spread and the ensemble mean RMSE -- called the spread-error ratio -- as a function of the lead time:
\begin{align}
\text{SER} = \frac{\text{StdDev}\left(\{x_{ens}\}_{i=1}^{ens}\right)}{\text{RMSE}\left(\langle{x_i \rangle}_{i=1}^{ens}, x_{target}\right)} \times \sqrt{\frac{ens + 1}{ens}}
\end{align}
where $\lbrace x_{ens} \rbrace_{i=1}^{ens}$ is an ensemble forecast with $ens$ denoting the number of ensemble members, $\langle \cdot \rangle$ indicates an ensemble mean and $x_{target}$ denotes the ground truth or target verification data.

A perfectly calibrated forecast will have a constant spread-error ratio of 1 indicating that the spread in the ensemble members is perfectly indicative of the uncertainty in the forecast.

Our results shown in Fig.~\ref{fig:spread-skill} indicate an under-dispersive ensemble. Future directions for improving the calibration of the ensemble could include using a GEFS ensemble as synoptic scale conditioning and incorporating initial condition uncertainty in the initialization of StormCast in addition to improving the forecast skill of StormCast.

\refstepcounter{AppendixCounter} 
\section*{Appendix \theAppendixCounter: Comparison with MPAS ensemble forecasts}\label{app:mpas}
\setcounter{figure}{0}

\begin{figure}[h]
    \centering
    \includegraphics[width=\textwidth]{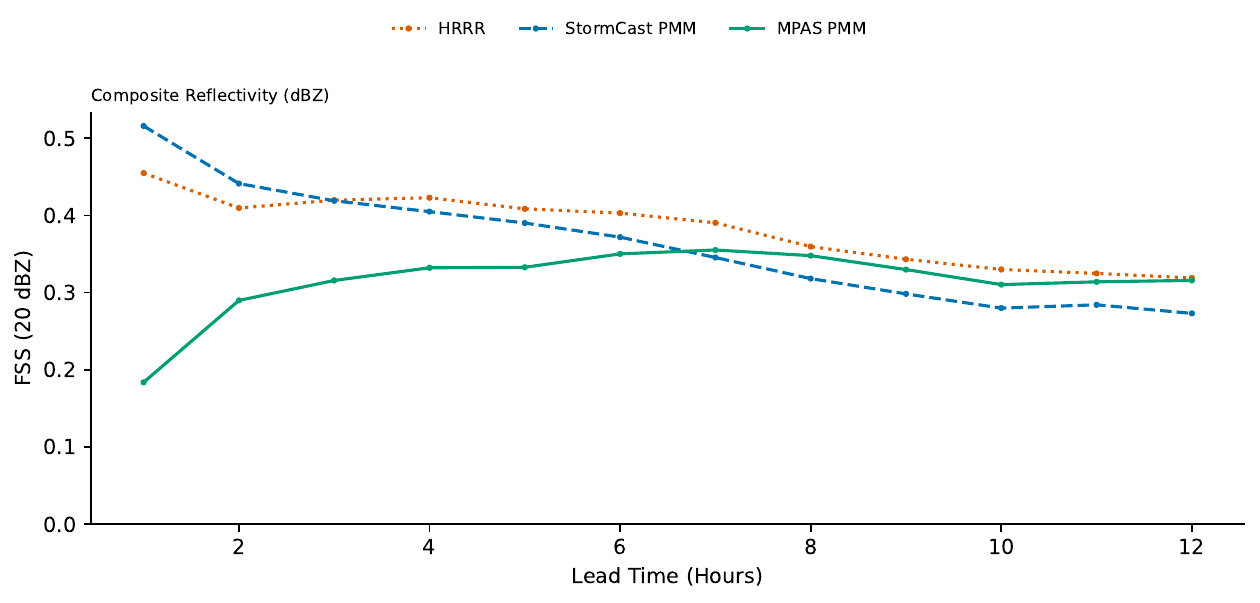}
    \caption{FSS at grid scale for a threshold of 20dBZ of StormCast 5-member ensemble forecast Probability Matched Mean (PMM), MPAS 5-member ensemble forecast PMM, and HRRR single member forecast. The FSS were averaged over a small set of 15 forecasts. Note that the MPAS forecasts were initialized from a GEFS interpolated state and did not use a radar-assimilated initial condition (HRRR analysis) which StormCast. The HRRR and StormCast forecasts have better skill in the first 6 hours than MPAS likely due to the difference in initialization techniques.}
    \label{fig:mpas}
\end{figure}

A small set of forecasts were obtained from the NSF NCAR MPAS Hazardous Weather Testbed (HWT) Spring Forecasting Experiment~\cite{NCAR_HWT2024, schwartz2024evaluation}. We found 15 forecasts where we had a corresponding StormCast forecast for the same data and time. Figure~\ref{fig:mpas} shows a comparison of the FSS of the StormCast 5-member ensemble forecast PMM, the MPAS 5-member ensmeble forecast PMM as well as a HRRR single member forecast. The MPAS forecasts were initialized from synoptic-scale GEFS analysis and did not assimilate radar data at initialization. This is likely why HRRR and StormCast forecasts have better skill in the first 6 hours. We hope to repeat this analysis with a radar-assimilating ensemble CAM forecast in future work.

\refstepcounter{AppendixCounter} 
\section*{Appendix \theAppendixCounter: Additional forecast case studies}\label{app:case_studies_supp}
\setcounter{figure}{0}

\begin{figure}[h]
    \centering
    \includegraphics[width=\textwidth]{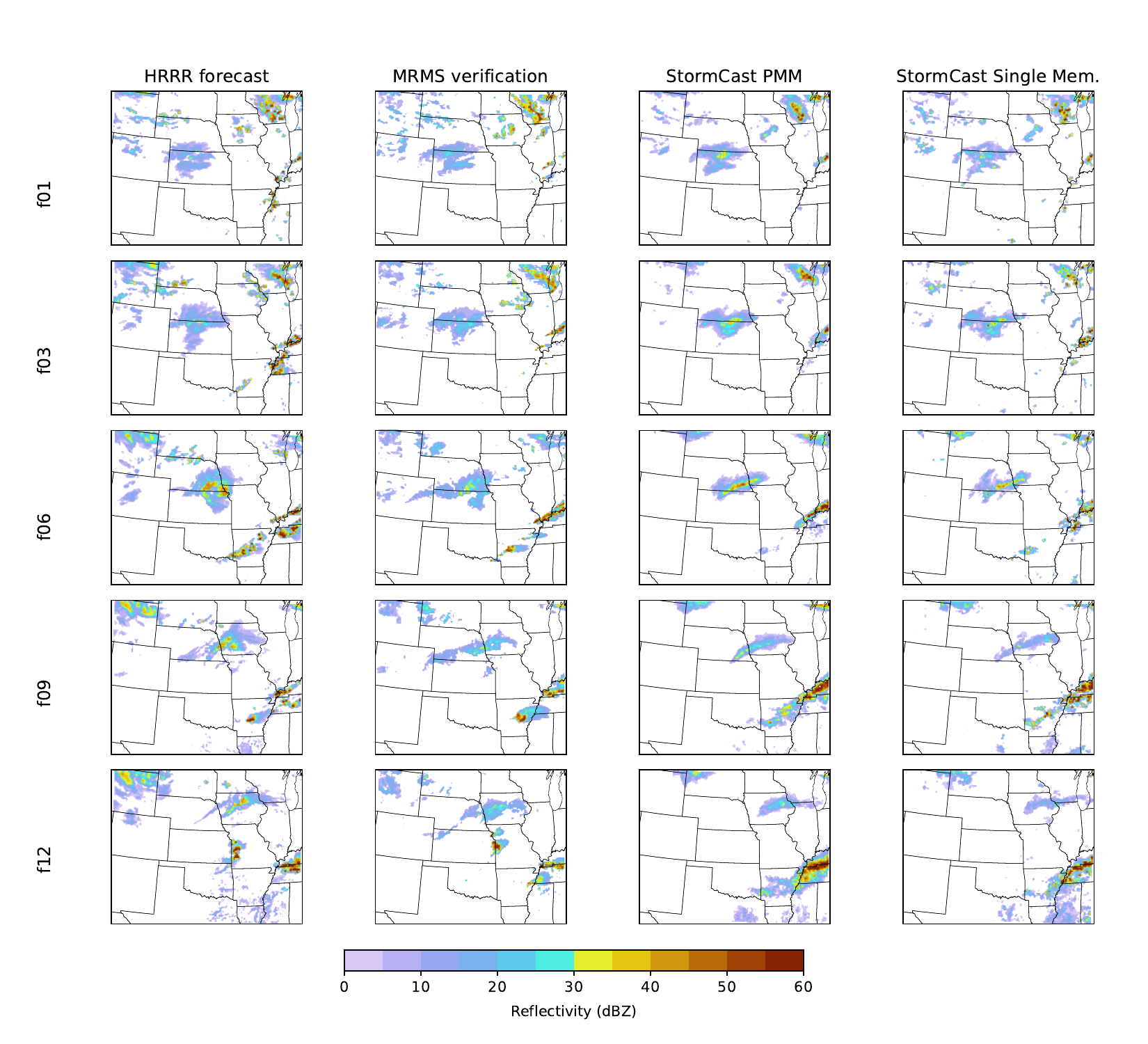}
    \caption{2024-05-08 00:00:00 initialization}
    \label{fig:cs1}
\end{figure}

\begin{figure}[h]
    \centering
    \includegraphics[width=\textwidth]{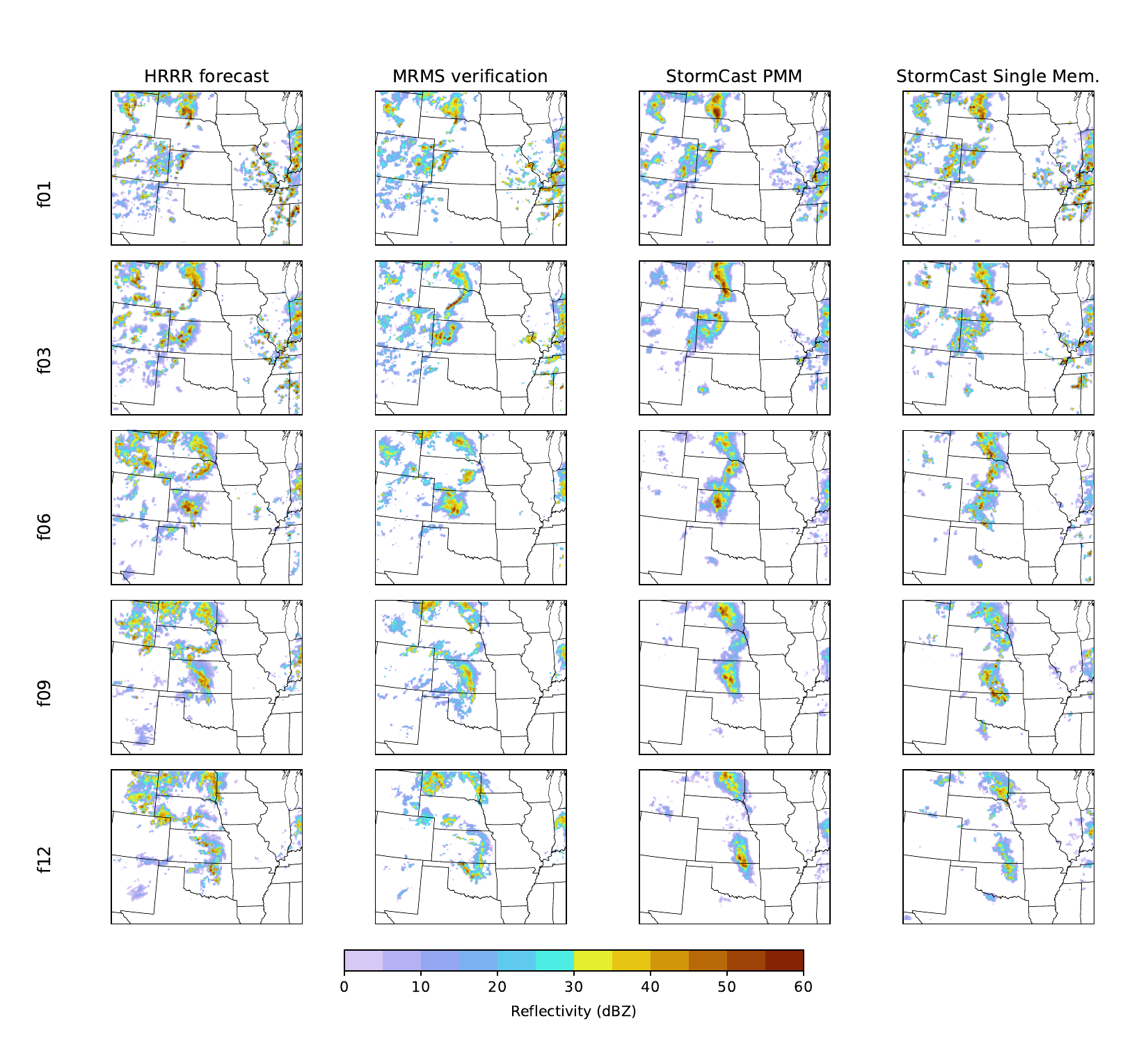}
    \caption{2024-05-15 00:00:00 initialization}
    \label{fig:cs2}
\end{figure}

\begin{figure}[h]
    \centering
    \includegraphics[width=\textwidth]{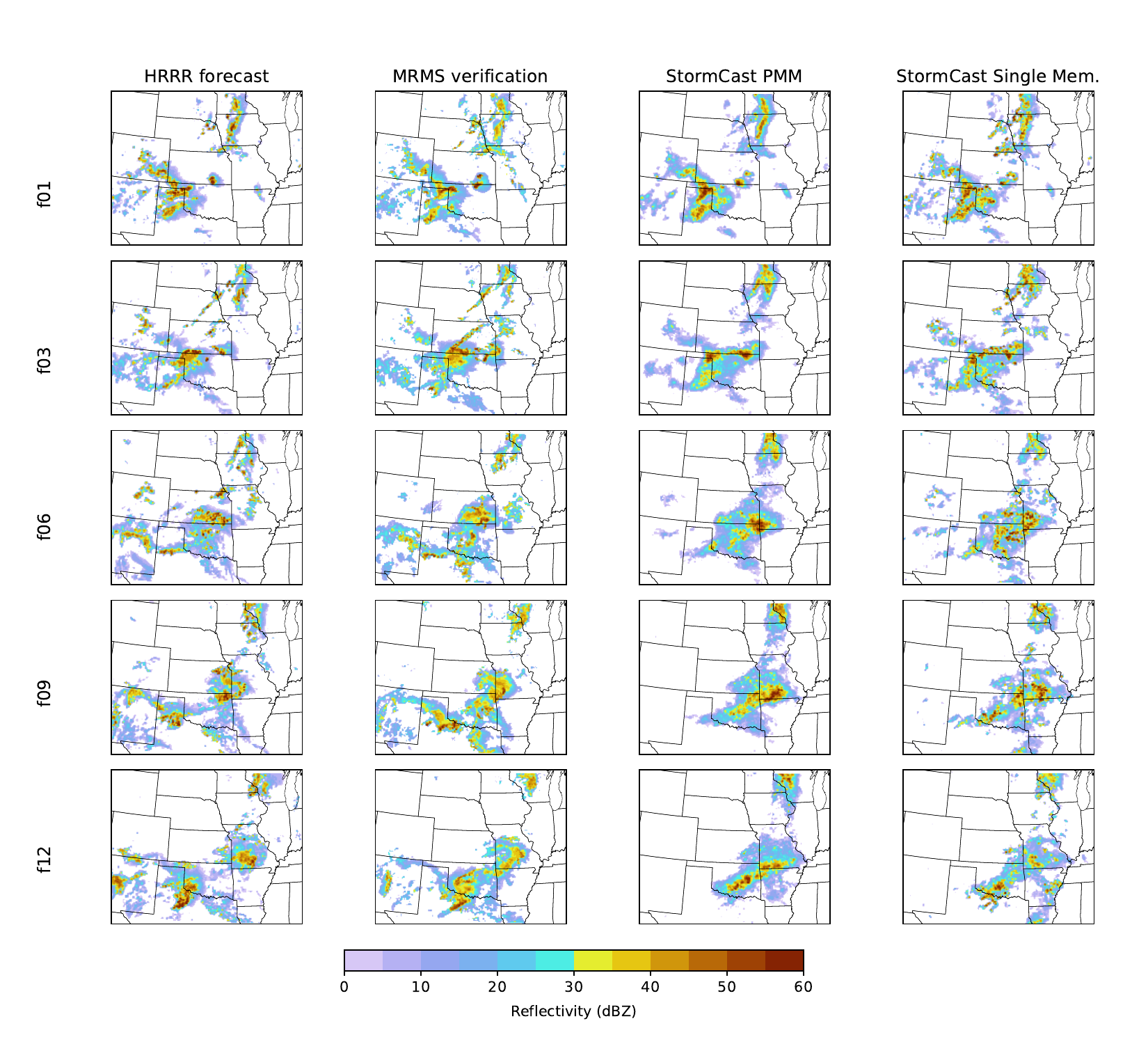}
    \caption{2024-05-16 00:00:00 initialization}
    \label{fig:cs3}
\end{figure}

\begin{figure}[h]
    \centering
    \includegraphics[width=\textwidth]{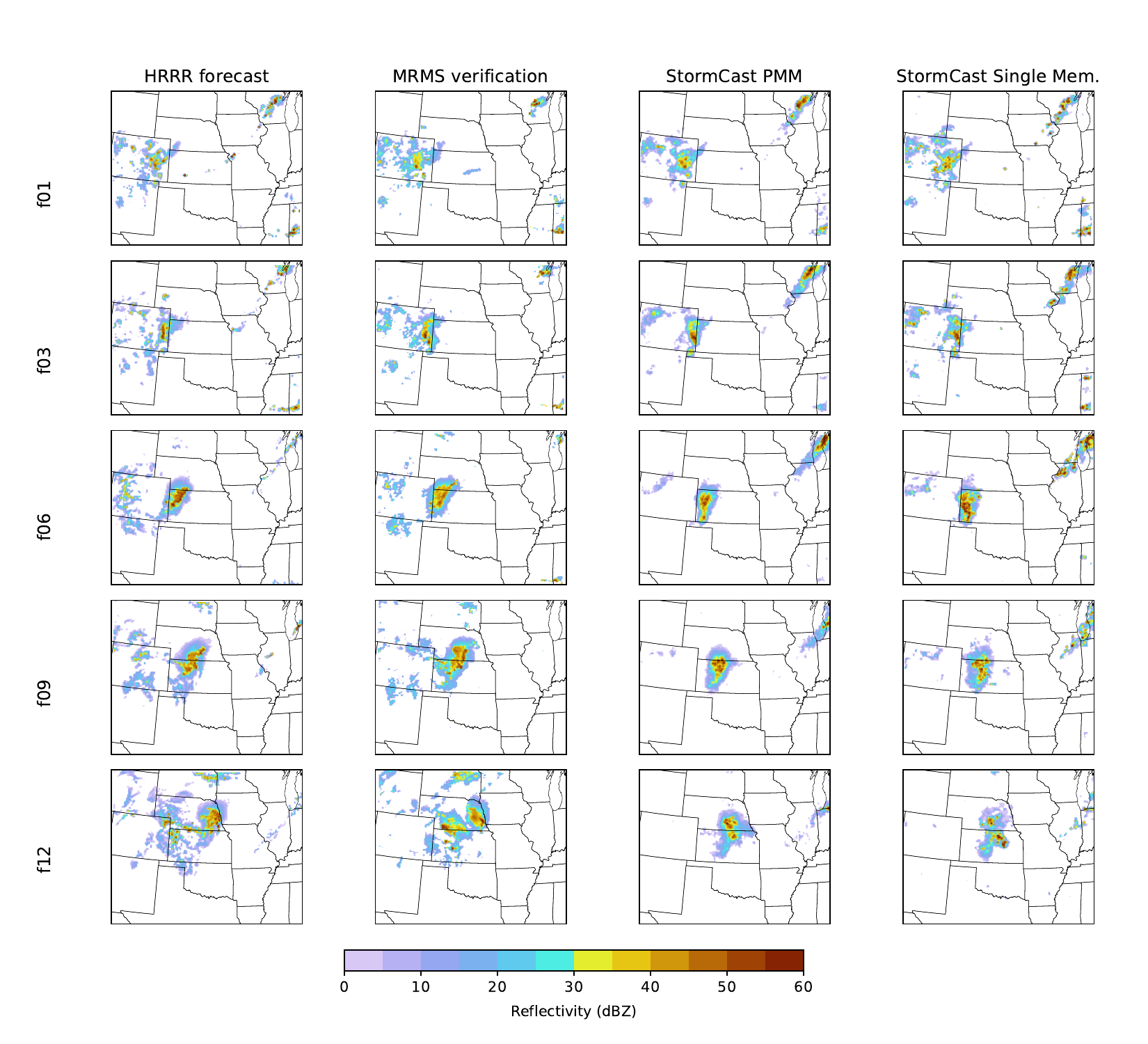}
    \caption{2024-05-19 00:00:00 initialization}
    \label{fig:cs4}
\end{figure}

\begin{figure}[h]
    \centering
    \includegraphics[width=\textwidth]{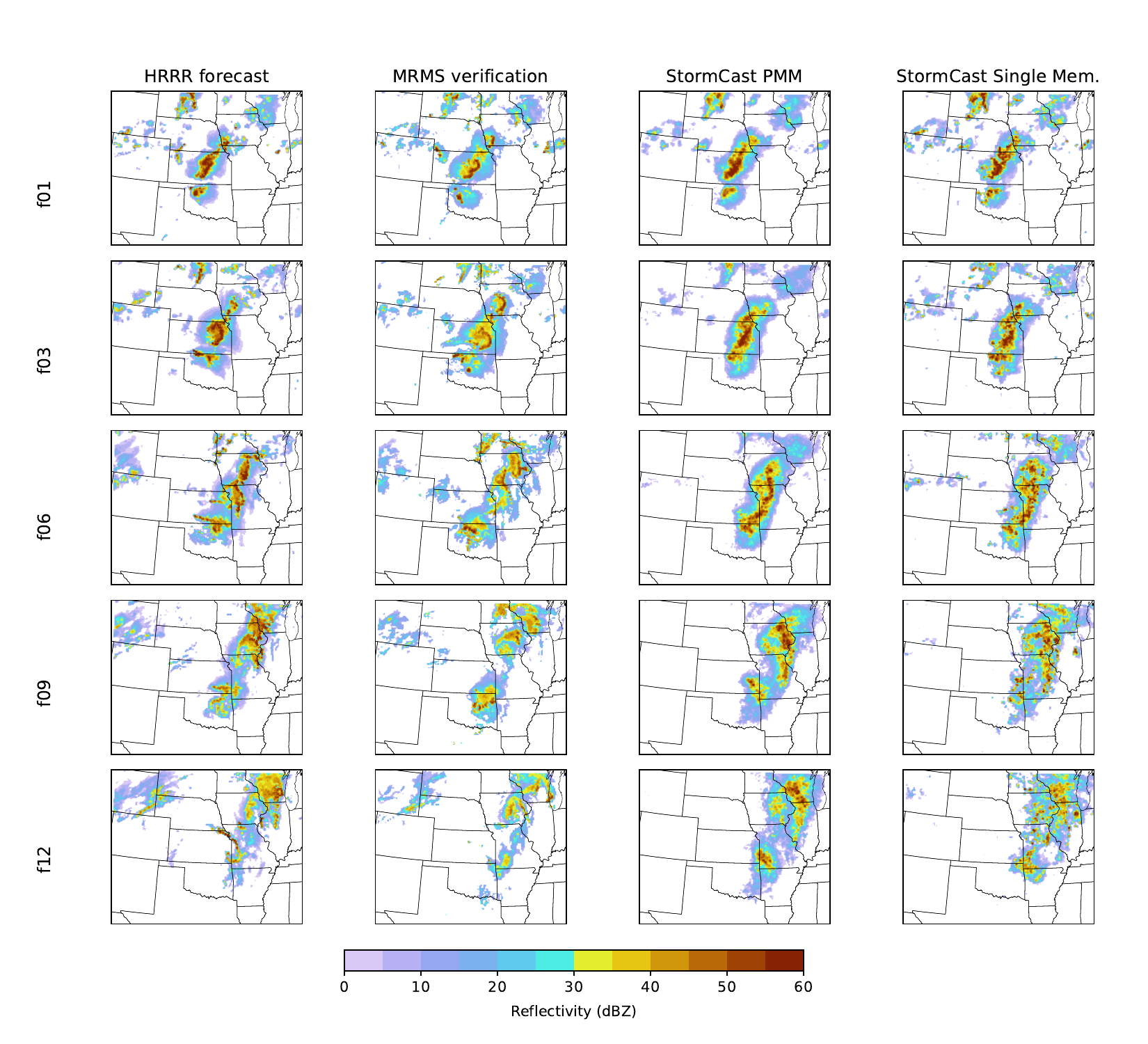}
    \caption{2024-05-20 00:00:00 initialization}
    \label{fig:cs5}
\end{figure}

\begin{figure}[h]
    \centering
    \includegraphics[width=\textwidth]{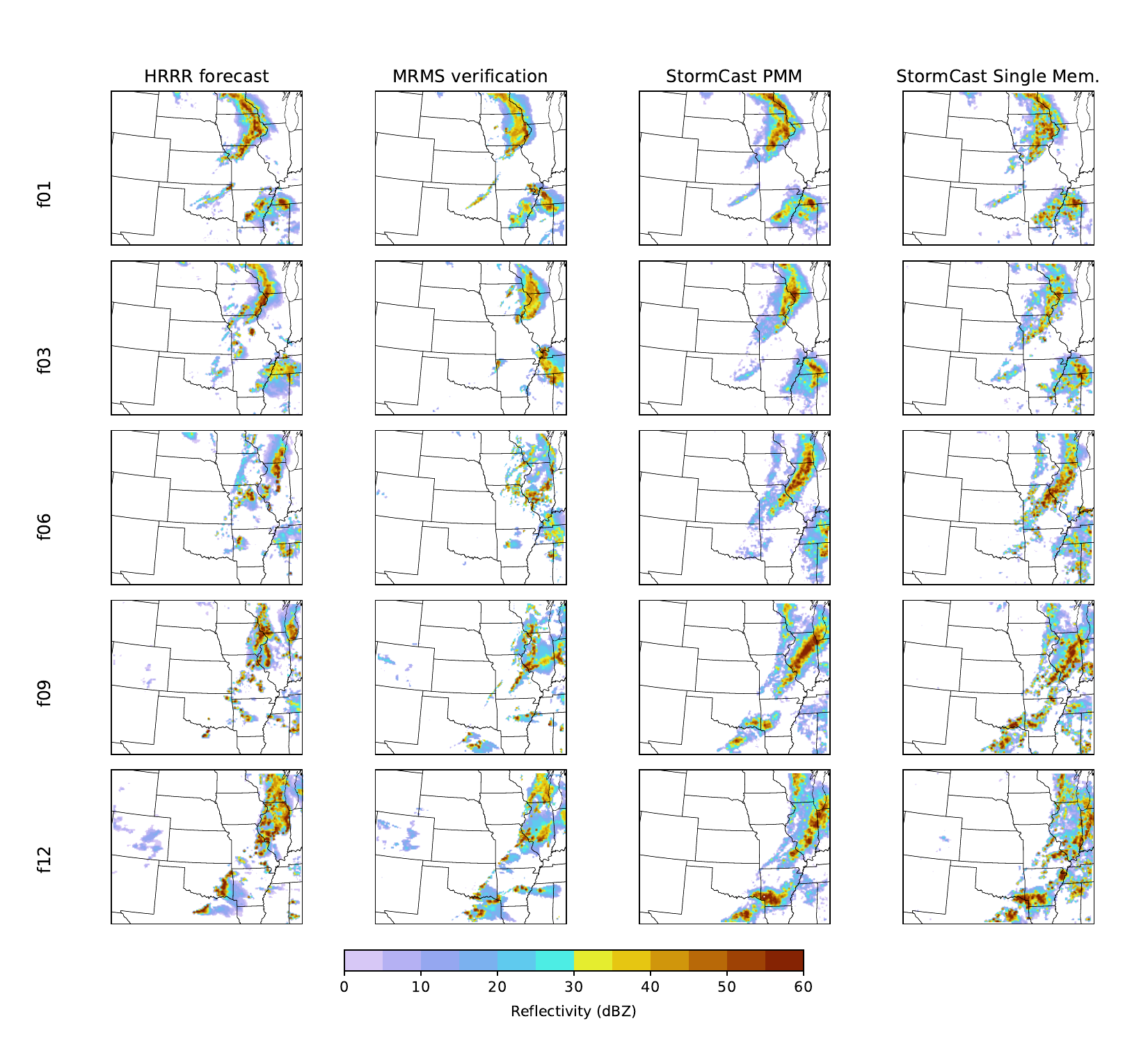}
    \caption{2024-05-24 12:00:00 initialization}
    \label{fig:cs6}
\end{figure}

\end{document}